\documentclass[preprint,aps,pra,showpacs,floatfix]{revtex4-1}

\usepackage[english]{babel}
\usepackage{amssymb}
\usepackage{amsmath,amsfonts,amssymb}
\usepackage{lscape}
\usepackage[dvips]{graphicx}
\usepackage[section]{placeins}
\usepackage{bm}

\usepackage{epsfig}

\newcommand{\bfr}{{\bf r}}

\newcommand{\balpha}{{\mbox{\boldmath$\alpha$}}}
\newcommand{\bnabla}{\bm{\nabla}}

\newcommand{\al}{\alpha}
\newcommand{\az}{\alpha Z}

\newcommand{\kk}{\lambda}

\newcommand{\albi}{\boldsymbol{\alpha}_i}
\newcommand{\albj}{\boldsymbol{\alpha}_j}

\newcommand{\la}{\langle}

\newcommand{\ra}{\rangle}
\newcommand{\be}{\begin{eqnarray}}
\newcommand{\ee}{\end{eqnarray}}
\newcommand{\pr}{\prime}
\newcommand{\veps}{\varepsilon}
\newcommand{\rmd}{{\rm d}}
\begin{document}
\title{Relativistic recoil, electron-correlation, and QED effects
on the $2p_j-2s$ transition energies in Li-like ions}
%
%
\author{Y.~S.~Kozhedub$^1$, A.~V.~Volotka$^{1,2}$, A.~N.~Artemyev$^3$,
D.~A.~Glazov$^1$,
G.~Plunien$^2$, V.~M.~Shabaev$^1$, I.~I.~Tupitsyn$^1$,  
and Th.~St\"ohlker$^{3,4}$}
\affiliation{
$^1$
Department of Physics, St. Petersburg State University,
Oulianovskaya 1, Petrodvorets, St. Petersburg 198504, Russia \\
$^2$
Institut f\"ur Theoretische Physik, Technische Universit\"at Dresden,
Mommsenstra{\ss}e 13, D-01062 Dresden, Germany \\
$^3$
Physikalisches Institut, Universit\"at Heidelberg, Philosophenweg 12, D-69120
Heidelberg, Germany \\
$^4$
GSI Helmholtzzentrum f\"ur Schwerionenforschung GmbH,
Planckstrasse 1, D-64291 Darmstadt, Germany \\
}
\begin{abstract}
The relativistic nuclear recoil, higher-order interelectronic-interaction,
and screened QED corrections to the transition energies in Li-like
ions are evaluated. The calculation of the relativistic recoil effect
is performed
to all orders in $1/Z$. The interelectronic-interaction correction
to the transition energies beyond the two-photon exchange level is evaluated
to all orders in $1/Z$ within the Breit approximation. The evaluation is carried
out employing the large-scale configuration-interaction Dirac-Fock-Sturm method.
The rigorous calculation of the complete gauge invariant sets of the screened
self-energy and vacuum-polarization diagrams is performed utilizing a local
screening potential as the zeroth-order approximation. The theoretical
predictions for the $2p_j-2s$ transition energies are compiled and compared with
available experimental data in the range of the nuclear charge number $Z = 10 -
60$.
\end{abstract}
\maketitle
%
%
\section{Introduction}
High-precision spectroscopy of Li-like ions continues to be of interest both
theoretically and experimentally. On the one hand such ions are among the
simplest few-electron systems that can be theoretically described with high
accuracy, on the other hand high precision measurements are also available.
Investigations of such systems enable precision tests of quantum
electrodynamics (QED) at strong fields, as well as studying various nuclear
properties probed by the atomic structure.
During the last decades significant theoretical efforts were undertaken to
evaluate various contributions to the energy levels in high-$Z$ Li-like ions
\cite{Blundell:93:1790,Lindgren:pra:1993,Chen:pra:1995,
yerokhin:1999:3522,
artemyev:1999:45,
PRL85_4699,
Zherebtsov:00:227,
Yerokhin:pra:2001,Sapirstein:2001:022502,Andreev:pra:2001,
Artemyev:pra:03,Yerokhin:prl:2006, Yerokhin:pra:2007,
Kozhedub:pra:2007,
Yerokhin:pra:2008,Kozhedub:pra:2008}.
However, further improvements in theoretical calculations are required in order
to meet the high level of the experimental accuracy
\cite{Edlen:ps:1983,Schweppe:prl:1991,Beiersdorfer:prl:1998,Feili:pra:2000,
Brandau:prl:2003,Beiersdorfer:prl:2005,Bushaw:pra:2007,Epp:prl:2007,
Lestinsky:prl:2008,Zhang:pra:2008}.

This work is devoted to high precision calculations of the $2p_j-2s$
transition
energies in middle-$Z$ Li-like ions. As was noticed in
Ref.~\cite{Yerokhin:pra:2007}, 
the leading sources of theoretical uncertainty originate from the relativistic 
recoil and higher-order screened QED corrections. Therefore, the present
paper is mainly focused on evaluation of these corrections.
The paper is organized as follows:
Sec.~\ref{MS_sec} is devoted to the calculation of the relativistic nuclear
recoil effect employing the large-scale configuration-interaction
Dirac-Fock-Sturm
method (CI-DFS). The method used for the calculation of the higher-order (in
$1/Z$)
relativistic recoil corrections allows us also to obtain accurate numerical
values for the interelectronic-interaction contributions to the transition
energies
within the Breit approximation. In Sec.~\ref{sec_HO} these results are combined
with the rigorous QED calculation of the one- and two-photon exchange
contributions to obtain the higher-order electron-electron
interaction
corrections
to the transition energies with the same accuracy level as in
Ref.~\cite{Yerokhin:pra:2007}. The calculation of the screened QED corrections
is presented in Sec.~\ref{ScrQED}. A local screening potential is included in
the
zeroth-order Hamiltonian. Then, the first and second-order diagrams representing
the screened self-energy (SE) and vacuum-polarization (VP) corrections are
rigorously evaluated.
In the last section, we compile all the contributions to
get the most accurate theoretical predictions for the $2p_{1/2}-2s$ and
$2p_{3/2}-2s$ transition energies of 
Li-like ions in the range of the nuclear charge number $Z = 10 - 60$ and compare
them with experimental data available.

Relativistic units ($\hbar=1$, $c=1$, $m=1$) and the Heaviside charge
unit [$\alpha = e^2 / (4\pi)$, $e<0$] are used throughout the paper.
\section{Relativistic theory of the nuclear recoil effect}
\label{MS_sec}
Since the electron mass is small compared to the nucleus mass, most of the
contributions to the binding energies can be evaluated within the infinite
nuclear mass approximation. Taking into account a finite nuclear mass shifts
the energies. This is so called nuclear recoil effect. Since this
effect is different for different isotopes, it also results in an isotope shift
of the energy levels. Generally, the isotope shift arises as a sum of the
finite nuclear mass effect (mass shift) and a non-zero nuclear size effect
(field shift). In this section we focus on calculations of the mass shift in
Li-like ions.

\subsection{Basic formulas}
In the nonrelativistic theory the mass shift (MS) is usually represented as a
sum of the normal mass shift (NMS) and the specific mass shift (SMS),
$H_{M}^{\rm (nonrel)}=H_{NMS}+H_{SMS}$, where \cite{Hughes:30:694}
\begin{align}
\label{nr_nms}
	H_{\rm NMS}&=\frac{1}{2M}\sum_i {\bf p}^{2}_i,\notag\\
	H_{\rm SMS}&=\frac{1}{2M}\sum_{i\ne j} {\bf p}_i\cdot{\bf p}_j.
\end{align}
Here, ${\bf p}_i$ is the electron momentum operator and $M$ is the nuclear mass.

A rigorous relativistic theory of the mass shift can be formulated only
in the framework of QED. Such a theory was formulated in
Refs.~\cite{Shabaev:TMP:1985,Shabaev:YP:1988} (see also
Refs.~\cite{Shabaev:98:57,Adkins:pra:2007} and references
therein), where the
complete
$\az$-dependent formulas for the recoil correction to the atomic energy
levels to first order in $m/M$ were derived.
Within the Breit approximation this theory leads to the
following many-body relativistic MS Hamiltonian:
\begin{align}
\label{r_ms}
	H_{M}=\frac{1}{2M}\sum_{i,j}\left\{ {\bf p}_i\cdot{\bf p}_j
	-\frac{\az}{r_i}\left[\albi+\frac{(\albi\cdot {\bf r}_i)\,{\bf r}_i}
	{r^{2}_{i}}\right]\cdot{\bf p}_j\right\},
\end{align}
where $\boldsymbol{\alpha}$ is a vector incorporating the Dirac matrices.
An  independent derivation of Hamiltonian
(\ref{r_ms}) was presented in Ref.~\cite{Palmer:87:5987}.
As follows from expression (\ref{r_ms}), the lowest-order
relativistic
correction to the one-electron mass shift operator is given by 
\begin{align}
\label{r_nms}
	H_{\rm RNMS}=-\frac{1}{2M}\sum_{i}
	\frac{\az}{r_i}\left[\albi+\frac{(\albi\cdot {\bf r}_i)\,{\bf r}_i}
	{r^{2}_{i}}\right]\cdot{\bf p}_i,
\end{align}
where ``RNMS'' stays for the relativistic NMS.
The corresponding two-electron correction is
\begin{align}
\label{r_sms}
	H_{\rm RSMS}=-\frac{1}{2M}\sum_{i\ne j}
	\frac{\az}{r_i}\left[\albi+\frac{(\albi\cdot {\bf r}_i)\,{\bf r}_i}
	{r^{2}_{i}}\right]\cdot{\bf p}_j,
\end{align}
where ``RSMS`` denotes the relativistic SMS.

The recoil correction to a given atomic state to first order in $m/M$ is
obtained as the expectation value of $H_M$ on the Dirac wave function (here and
in what follows, the Dirac wave functions are the eigenvectors of the
Dirac-Coulomb-Breit Hamiltonian). 
In Ref.~\cite{Shabaev:94:l307} the Hamiltonian (\ref{r_ms}) was employed to
calculate the
$(\az)^4 m/M$ corrections to the energy levels in He- and Li-like
ions to zeroth order in $1/Z$. Later in
Refs.~\cite{Tupitsyn:03:022511,korol:pra:07},
this Hamiltonian was used to evaluate the relativistic recoil effect in low-
and middle-$Z$ ions and atoms to all orders in $1/Z$.

The recoil correction of the first order in $m/M$ is conveniently expressed in
terms of the constant $K$ defined by
\begin{align}
\label{K_ms_r}
\Delta E =\la\psi|H_{M}|\psi\ra \equiv K/M,
\end{align}
where $|\psi\ra$ is the
eigenvector of the Dirac-Coulomb-Breit Hamiltonian.
With this constant, the mass isotope shift for two different isotopes with
nuclear masses $M_1$ and $M_2$ can be written as 
$\delta E=K\left(\frac{1}{M_1}-\frac{1}{M_2}\right)$.

The recoil correction which is beyond the Breit approximation (\ref{r_ms})
is referred to a QED recoil effect. This effect has to be also taken into
account,
especially for high-$Z$ ions. For H- and Li-like ions the QED
recoil corrections have been calculated to all orders in $\az$ and to zeroth
order in $1/Z$ in Refs.~\cite{Artemyev:pra:95,Artemyev:jpb:95}.
In what follows, we focus on the calculations of the coefficient $K$
to all orders
in $1/Z$  for the $2p_j-2s$ transitions in a wide range of Li-like ions.
We investigate relative contributions of the relativistic and QED corrections
to the total recoil effect and the influence of the electron correlations on
the recoil effect.
\subsection{Method of calculation}
Expectation values of the MS operator (\ref{r_ms}) are very sensitive
to the electron correlations. In the present
investigation the large-scale configuration-interaction (CI) Dirac-Fock-Sturm
(DFS) method was employed to solve the Dirac-Coulomb-Breit equation with high
accuracy. 
This method was developed by Tupitsyn and partially presented
in Ref.~\cite{Tupitsyn:pra:2005}. It was successfully used for 
calculations of the recoil effect in
Refs.~\cite{Tupitsyn:03:022511,Kozhedub:pra:2007}.
The MS is calculated as the expectation value of the recoil
operator with the many-electron Dirac wave function.
Additionally, we apply an alternative approach which consists in adding
the 
operator $H_{M}$
(\ref{r_ms}) to the many-electron Hamiltonian $H$ with an arbitrary
coefficient $\lambda$
\begin{equation}
\label{H_lambda}
H(\lambda) = H + \lambda H_{M}
\end{equation}
and evaluating the MS by
\begin{equation}
\label{derivative}
  \Delta E=\frac{\rmd}{\rmd\kk}E(\lambda)\Big|_{\lambda=0}.
\end{equation}
Here the derivative is determined numerically and $\lambda$
is chosen obeying the numerical stability and smallness of the nonlinear terms.
We have reformulated the CI-DFS method to adopt the alternative scheme and
independently evaluated the normal and specific parts of the MS
by both methods.
\subsection{Results of the calculations and discussion}
\begin{table}[tub]
\caption{Individual contributions to the mass shift coefficient $K$
(GHz$\cdot$amu) for the $2p_{1/2}-2s$ and $2p_{3/2}-2s$ transitions in lithium
(Z=3).
}
\label{IS_table1}
\begin{center}
\begin{tabular}{llllr}
 \hline
 \hline
&Subset&\multicolumn{1}{c}{ $ 2p_{1/2}-2s$}
&\multicolumn{1}{c}{ $ 2p_{3/2}-2s$}&Ref. \\ \hline
    MS operator& &      -443.81(20)&      -443.82(20)&\\                 
    & &      -443.86$^{\rm{nr}}$&      -443.86$^{\rm{nr}}$&\\
    & &     -2534.48$^{\rm{hyd}}$&     -2535.12$^{\rm{hyd}}$&\\
    &NMS &      -245.48&      -245.49&\\
    &SMS &      -198.78&      -198.77&\\
    & &      -198.73$^{\rm{nr}}$&      -198.73$^{\rm{nr}}$&\\
    & &      -198.920(2)$^{\rm{nr}}$&     
-198.920(2)$^{\rm{nr}}$&\cite{Luchow:92:105,Luchow:94:211,Luchow:97:61}\\
    & &      -198.8$^{\rm{nr}}$&     
-198.8$^{\rm{nr}}$&\cite{Godefroid:01:1079}\\
    &RNMS &         0.33&         0.38&\\
    &RSMS &         0.12&         0.06&\\
    QED& &        -0.08(3)&        -0.08(3)&\\
    &1-el QED &        -0.08&        -0.08&\\
    &2-el QED &         0.00&        0.00&\\
    Total theory& &      -443.9(2)&      -443.9(2)&\\
          & &      -444.086&     
-444.103&\cite{Yan:prl+erratum:2008,Puchalski:pra:2008}\\
          & &      -447(12)&      -447(12)&\cite{korol:pra:07}\\
    Experiment$^{\ast}$   & &      -444.09(3)&             
&\cite{Walls:epjd:2003}\\
                    & &      -444.04(4)&   
-444.06(4)&\cite{Sansonetti:pra:1995}\\
 \hline
 \hline
 \multicolumn{5}{l}{$^{\ast}$ The experimental values include also
terms of higher orders in $m/M$.}\\
\end{tabular}
\end{center}
\end{table}
\begin{table}[tub]
\caption{Individual contributions to the mass shift coefficient $K$
(GHz$\cdot$amu) for the $2p_{1/2}-2s$ and $2p_{3/2}-2s$ transitions in Li-like zinc
(Z=30).
}
\label{IS_table2}
\begin{center}
\begin{tabular}{llllr}
 \hline
 \hline
&Subset&\multicolumn{1}{c}{ $ 2p_{1/2}-2s$}
&\multicolumn{1}{c}{ $ 2p_{3/2}-2s$}&Ref. \\ \hline
    MS operator& &   -224600(3)&   -230073(3)&\\
    & &   -230161$^{\rm{nr}}$&   -230161$^{\rm{nr}}$&\\
    & &   -246954$^{\rm{hyd}}$&   -253951$^{\rm{hyd}}$&\\
    &NMS &    -21862.0&    -34139.7&\\
    &SMS &   -235922.0&   -225509.0&\\
    &RNMS &     13807.8&     22890.0\\
    &RSMS &     19377.1&      6685.8&\\
    QED& &     -591(20)$\times10$&     -560(20)$\times10$&\\
    &1-el QED &     -5411&     -5504&\\
    &2-el QED &      -497&       -98\\
    Total theory& &   -23051(20)$\times10$&   -23568(20)$\times10$&\\
 \hline
 \hline
\end{tabular}
\end{center}
\end{table}
\begin{table}[tub]
\caption{Individual contributions to the mass shift coefficient $K$
(GHz$\cdot$amu) for the $2p_{1/2}-2s$ and $2p_{3/2}-2s$ transitions in Li-like
neodymium (Z=60).
}
\label{IS_table3}
\begin{center}
\begin{tabular}{llllr}
 \hline
 \hline
&Subset&\multicolumn{1}{c}{ $ 2p_{1/2}-2s$}
&\multicolumn{1}{c}{ $ 2p_{3/2}-2s$}&Ref. \\ \hline
    MS operator& &   -834508(25)&   -962662(25)&\\
    & &   -967156$^{\rm{nr}}$&   -967156$^{\rm{nr}}$&\\
    & &   -868746$^{\rm{hyd}}$&  -1014056$^{\rm{hyd}}$&\\
    &NMS &   -339895&   -625227&\\
    &SMS &  -1227059&   -998772&\\
    &RNMS &    323477&    539494&\\
    &RSMS &    408969&    121843&\\
    QED& &   -2133(35)$\times10^2$&   -1958(35)$\times10^2$&\\
    &1-el QED &   -175010&   -188435&\\
    &2-el QED &    -38285&     -7332&\\
    Total theory& &  -10478(35)$\times10^2$&-11584(35)$\times10^2$&\\
 \hline
 \hline
\end{tabular}
\end{center}
\end{table}
\begin{table}[tub]
\caption{Individual contributions to the mass shift coefficient $K$
(1000 GHz$\cdot$amu) for the $2p_{1/2}-2s$ and $2p_{3/2}-2s$ transitions in Li-like
uranium (Z=92).
}
\label{IS_table4}
\begin{center}
\begin{tabular}{llllr}
 \hline
 \hline
&Subset&\multicolumn{1}{c}{ $ 2p_{1/2}-2s$}
&\multicolumn{1}{c}{ $ 2p_{3/2}-2s$}&Ref. \\ \hline
    MS operator& &   -733&  -2010&\\
    & &  -2312$^{\rm{nr}}$&  -2312$^{\rm{nr}}$&\\
    & &   -768$^{\rm{hyd}}$&  -2174$^{\rm{hyd}}$&\\
    &NMS &  -3665&  -6671&\\
    &SMS &  -4633&  -2547&\\
    &RNMS &   3892&   6443&\\
    &RSMS &   3673&    764&\\
    QED& &  -3000(32)&  -2851(32)&\\
    &1-el QED &  -2222&  -2729&\\
    &2-el QED &   -778&   -122&\\
    Total theory& &  -3734(32)&  -4861(32)&\\
 \hline
 \hline
\end{tabular}
\end{center}
\end{table}

Here we examine our calculations of the mass shift coefficient $K$  in Li-like
ions and
compare them with the related results obtained by other authors. In
Tables \ref{IS_table1}, \ref{IS_table2}, \ref{IS_table3}, and \ref{IS_table4}
we present
numerical results for the coefficient $K$ calculated for the
$2p_{1/2}-2s$ and $2p_{3/2}-2s$ transitions in lithium, Li-like zinc, neodymium,
and uranium, respectively.
The first line shows the contribution obtained employing the MS
operator (\ref{r_ms}). The entries labeled ``NMS'', ``SMS'', ``RNMS'' and
``RSMS''
represent the corresponding contributions of the mass shift operators. Since the
expectation
values of the NMS and SMS operators are evaluated with the Dirac wave functions,
the values denoted by NMS and SMS in the tables partly contain the relativistic
contributions. The values marked by ``nr'' show the nonrelativistic values of
the corresponding
contributions, obtained within the same computing procedure but with a
1000-times increased
value of the speed of light (in atomic units). We have verified this
nonrelativistic
limit by comparing our values with the results of the fully nonrelativistic
method based
on the Schr\"{o}dinger Hamiltonian and on the same calculation scheme.
The values have exactly coincided with each other for all the ions under
consideration.
To demonstrate the importance of the electron-electron interaction
effects we present also the related results obtained with the
hydrogenlike wave functions. These values are marked as ``hyd'' in the tables.
Obviously, the CI-DFS approach
is not the best for very low-$Z$ Li-like systems. Most accurate results for
lithium are presently obtained  utilizing variational solutions of the
three-body Schr\"{o}dinger problem and accounting for the relativistic and QED
corrections within the $\az$ expansion
\cite{Yan:prl+erratum:2008,Puchalski:pra:2008}.
We use these results to estimate the residual correlation effects in
our
calculations. Analyzing the convergence of the calculated atomic properties
as a function of the configuration basis set, the difference between
the results
obtained
by two alternative methods described above, and the deviation of our
nonrelativistic SMS values from the related results by other nonrelativistic
calculations,
we estimate an uncertainty associated with the electron correlation as
$0.05\%$ for lithium, $0.002\%$ for Li-like boron and much less
for ions with larger nuclear charge numbers. 

One-electron and two-electron QED recoil corrections were calculated in
accordance with
our previous works \cite{Artemyev:pra:95,Artemyev:jpb:95,Shabaev:98:4235}. The
evaluation is performed for extended nuclei within the approximation of
noninteracting electrons.
The electron-electron interaction is suppressed by a factor $1/Z$, therefore we
estimate the uncertainty of the QED recoil contribution multiplying it by
$1/Z$.

As one can see, in the case of Li our values agree well with the previous
theoretical predictions
\cite{korol:pra:07,Luchow:92:105,Luchow:94:211,Luchow:97:61,Godefroid:01:1079,
Yan:prl+erratum:2008,Puchalski:pra:2008} as
well as with the experimental data
\cite{Walls:epjd:2003,Sansonetti:pra:1995}.

\begin{figure}[tub]
\begin{center}
\includegraphics[angle=0, width=12cm,height=9cm]{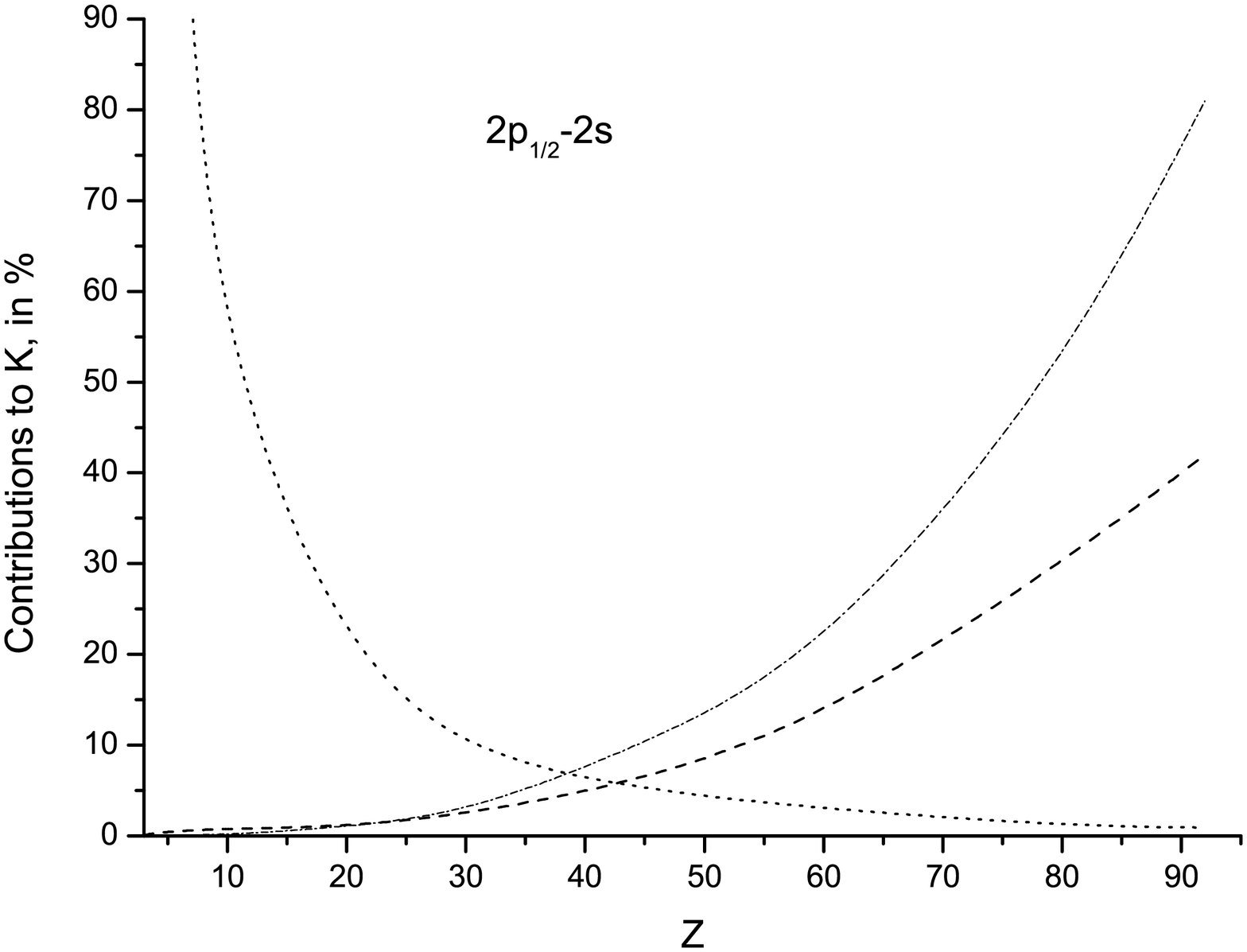}
\includegraphics[angle=0, width=12cm,height=9cm]{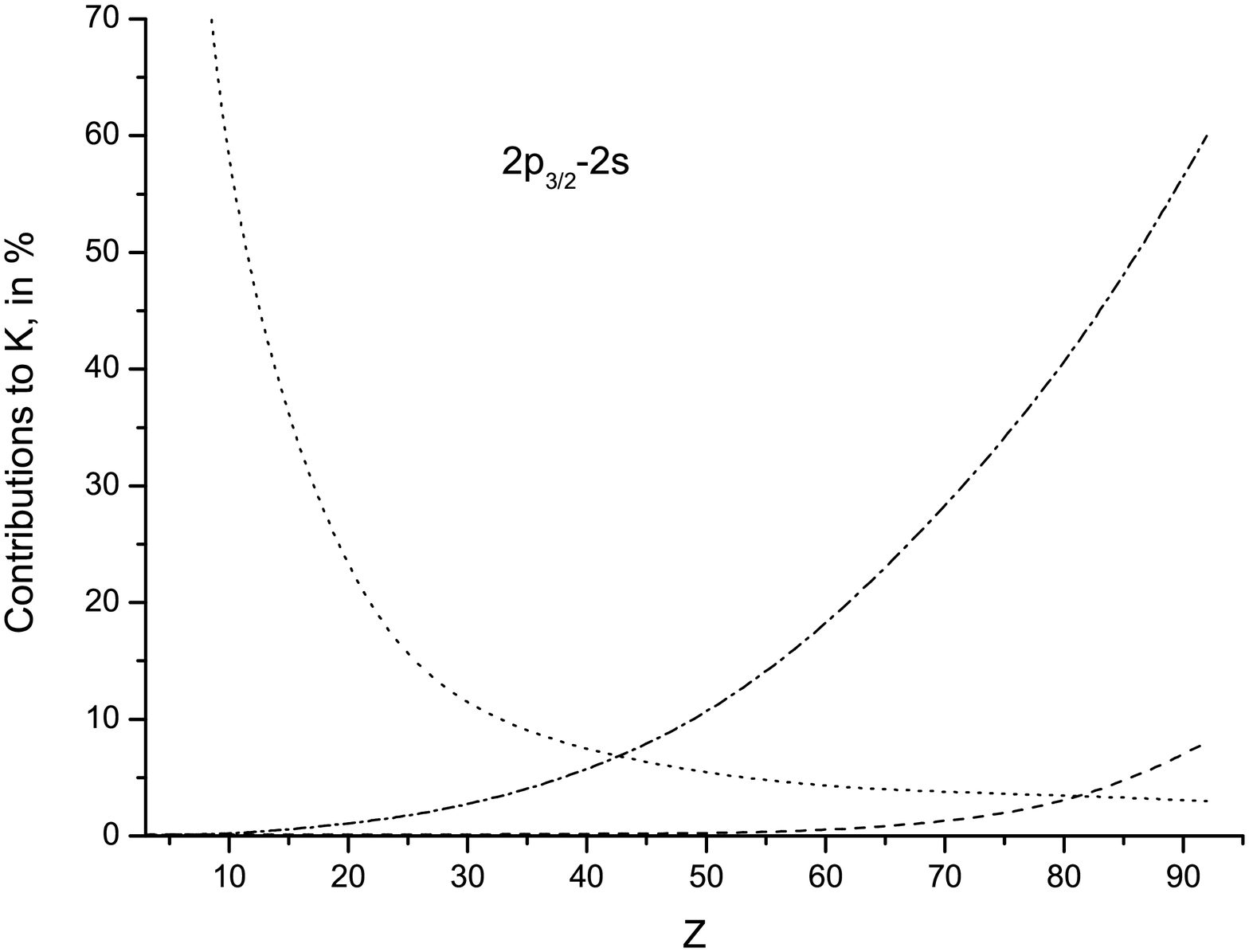}
\end{center}
\caption{Relative values (in $\%$) of the individual contributions to the mass
shift coefficient $K$ for
the $2p_{1/2}-2s$ and $2p_{3/2}-2s$ transitions in Li-like ions. The dotted
line represents the relative contribution of the electron-electron interaction;
the dashed line denotes the relativistic correction; and the dashed-dotted
line indicates the QED correction.}
\label{fig1}
\end{figure}
In Fig.~\ref{fig1} we plot the individual contributions to the MS
coefficient $K$ for the $2p_{1/2}-2s$ and $2p_{3/2}-2s$ transitions in Li-like
ions. The dotted line indicates the relative contribution of the electron-electron
interaction; the dashed line represents the relativistic correction; and the
dashed-dotted line stands for the QED part of the coefficient.  We observe that
for low-$Z$ ions it is extremely important to include the electron-electron
interaction effects. For middle-$Z$ ions all parts are equally important.
For high-$Z$ region the QED and relativistic contributions become dominant.
It is interesting to note that for the high $Z$ the QED contribution is larger
than the relativistic one. One can see also that the relativistic contribution
for the
$2p_{3/2}-2s$ transition is much smaller than for the $2p_{1/2}-2s$ one.
This is due to a large cancellation of the relativistic
NMS and relativistic SMS contributions for the
$2p_{3/2}-2s$ transition. We note also that for the $2p_{1/2}-2s$ transition
the NMS equals to zero in hydrogenlike ions
with a pointlike nucleus.

The total results for the MS coefficient $K$ for the $2p_{1/2}-2s$ and
$2p_{3/2}-2s$ transitions in Li-like ions with the nuclear
charge numbers $Z=3-92$ are presented in Table~\ref{IS_fin_table}.
Now the leading theoretical uncertainty for
middle- and high-$Z$ ions is determined by uncalculated electron-electron
interaction effects of the QED recoil contribution.

\begin{table}
\caption{Mass shift coefficient $K$ (GHz$\cdot$amu) for the $2p_{1/2}-2s$ and
$2p_{3/2}-2s$ transitions in Li-like ions.
}
\label{IS_fin_table}
\begin{center}
\begin{tabular}{lrrrrrr}
 \hline
 \hline
&\multicolumn{3}{c}{ $ 2p_{1/2}-2s$}
&\multicolumn{3}{c}{ $ 2p_{3/2}-2s$} \\
$Z$
&  \multicolumn{1}{c}{MS operator} & \multicolumn{1}{c}{QED}&
\multicolumn{1}{c}{Total} &
   \multicolumn{1}{c}{MS operator} & \multicolumn{1}{c}{QED}&
\multicolumn{1}{c}{Total}\\ \hline
      3&      -443.8(2)&        -0.08(3)&    -443.9(2)&      -443.8(2)&       
-0.08(3)&      -443.9(2)\\
      5&     -3281.4(5)&        -0.89(18)&   -3282.3(5)&     -3282.2(5)&       
-0.92(18)&     -3283.1(5)\\
     10&    -20420.0(5)&       -27.4(3.0)&  -20447(3)&    -20456.6(5)&      
-26.7(3.0)&    -20483(3)\\
     20&    -96182(2)&      -802(40)&  -9698(4)$\times10$&    -97092(2)&     
-774(40)&    -9787(4)$\times10$\\
30&   -22460.0(3)$\times10$&     -591(20)$\times10$& -23051(20)$\times10$&  
-23007.3(3)$\times10$&
-560(20)$\times10$&   -23568(20)$\times10$\\
40&   -3997.4(1)$\times10^2$&    -251(7)$\times10^2$& -4248(7)$\times10^2$&  
-4194(7)$\times10^2$&   
-234.1(2)$\times10^2$&   -4428(7)$\times10^2$\\
50&   -6102.5(2)$\times10^2$&    -796(16)$\times10^2$& -6899(16)$\times10^2$&
 
-6643.7(2)$\times10^2$&-735(16)$\times10^2$&   -7379(16)$\times10^2$\\
60&   -8345.1(3)$\times10^2$&   -2133(35)$\times10^2$&-10478(35)$\times10^2$&
 
-9626.6(3)$\times10^2$&  
-1958(35)$\times10^2$&  -11584(35)$\times10^2$\\
70&  -1029.41(4)$\times10^3$&   -515(7)$\times10^3$&-1544(7)$\times10^3$& 
-1305.85(4)$\times10^3$&  
-473(7)$\times10^3$&  -1779(7)$\times10^3$\\
80&  -1105.6$\times10^3$& 
-1167(14)$\times10^3$&-2272(14)$\times10^3$& 
-1669.5$\times10^3$& -1082(14)$\times10^3$&  -2751(14)$\times10^3$\\
92&   -733$\times10^3$&  -3000(32)$\times10^3$&-3734(32)$\times10^3$& 
-2010$\times10^3$& -2851(32)$\times10^3$&  -4861(32)$\times10^3$\\
\hline
 \hline
\end{tabular}
\end{center}
\end{table}

\section{Higher-order electron-correlation corrections to the
 transition energies}
\label{sec_HO}

Electron-electron interaction within the basic principles of QED
is described by exchange of virtual photons. The one-photon exchange leads to
the operator
\begin{equation}
\label{I:def}
I(\omega )=e^2\alpha^{\mu}_1\alpha^{\nu}_2 D_{\mu\nu}(\omega
,\boldsymbol{r}_{12}),
\end{equation}  
where $D_{\mu\nu}$ is the photon propagator, which in the Coulomb gauge is
written as 
\begin{align}
\label{prop_C}
D_{00}(\omega ,\boldsymbol{r}_{12})=\frac{1}{4\pi r_{12}},\quad 
D_{i0}=D_{0i}=0 \quad  (i=1,2,3)\,,\notag\\
D_{il}(\omega ,\boldsymbol{r}_{12})=\int
\frac{d\boldsymbol{k}}
{(2\pi)^3}\frac{\exp(i\boldsymbol{k}\cdot\boldsymbol{r}_{12})}{
\omega^2-\boldsymbol{k}^2+i0} 
\Big(
\delta_{il}-\frac{k_i k_l}{\boldsymbol{k}^2}
\Big) \quad  (i,l=1,2,3)\,,
\end{align}  
$r_{12}=|\boldsymbol{r}_{12}|=|\boldsymbol{r}_{1}-\boldsymbol{r}_{2}|$,
$\boldsymbol{r}_i$ is the position vector of the $i$th electron,
 and $\alpha^{\mu}=(1,\boldsymbol{\al})$ are the Dirac matrices. 

Expanding expression (\ref{prop_C}) in powers of the photon frequency one can
derive a simplified form of the interaction. The low-frequency limit of this
interaction consists of two parts, referred to
as the Coulomb and the Breit interaction,
\begin{equation}
\label{a1}
V(i,j)=V_{\rm{C}}(i,j)+V_{\rm{B}}(i,j)=\frac{\al}{r_{ij}} -
\al\left[\frac{\albi \cdot
\albj}{2r_{ij}} 
 + \frac{(\albi \cdot \bfr_{ij}) (\albj \cdot
\bfr_{ij})}{2r^{3}_{ij}}\right].
\end{equation}  
The most traditional approach for the treatment of
the electron-electron interaction in relativistic many-electron atoms consists
in using so-called Breit approximation. In this approximation the total
Hamiltonian can be represented as the sum of the one-electron Dirac
Hamiltonians and the Coulomb and Breit electron-electron interactions, projected
on the positive-energy Dirac's states. In this way one gets the
Dirac-Coulomb-Breit equation. 
Traditional methods for solving the Dirac-Coulomb-Breit equation
are the many-body perturbation theory (MBPT)
\cite{Johnson:88:2764,ynnerman:1994:4671},
the multi-configuration Dirac-Fock method \cite{indelicato:1990:5139}, and the
configuration-interaction (CI) method
\cite{Chen:pra:1995,Tupitsyn:03:022511}. All these
methods treat the one-photon exchange exactly and the
higher-order electron correlation is accounted for within the Breit
approximation only.

The current level of experimental accuracy demands rigorous QED calculations of
two-photon
exchange contributions, which for $n=2$ states of Li-like ions were
performed in
Refs.~\cite{PRL85_4699,Yerokhin:pra:2001,Sapirstein:2001:022502,
Andreev:pra:2001,Andreev:pra:2003,Artemyev:pra:03,Yerokhin:pra:2007}.
Meanwhile rigorous QED calculations of three- and more photon exchange
contributions
have not been performed up to now. For high-$Z$ few-electron ions
evaluations of these contributions within the Breit approximation
are generally sufficient. Previously such calculations for Li-like ions were
performed in Refs.~\cite{Zherebtsov:00:227,Andreev:pra:2001,
Sapirstein:2001:022502,Yerokhin:pra:2007}.
The evaluations of
Refs.~\cite{Zherebtsov:00:227,Andreev:pra:2001}
were carried out with the hydrogenic wave functions while in
Refs.~\cite{Sapirstein:2001:022502,Yerokhin:pra:2007,Artemyev:prl:2007}
the perturbation expansion starts with a local screening potential, which partly
incorporates the electron-electron interaction effects.

In the present investigation, to evaluate the
interelectronic-interaction corrections of the third and higher orders we
proceed as follows. The large-scale CI-DFS method (see, e.g.,
Refs.~\cite{Tupitsyn:03:022511,Tupitsyn:pra:2005}) was used to
solve the Dirac-Coulomb-Breit equation yielding the energies. The operator of
the interelectronic interaction in the Breit approximation reads
 \begin{eqnarray}
 \label{interaction}
  V(\kk) = \kk\al \sum_{i>j} \left[ \frac{1}{r_{ij}} - \frac{\albi
\cdot \albj}{2r_{ij}}
  - \frac{(\albi \cdot \bfr_{ij}) (\albj \cdot
\bfr_{ij})}{2r^{3}_{ij}}\right]
\,,
\end{eqnarray}
where a scaling parameter $\kk$ is introduced to separate
terms of different order in $1/Z$ using the numerical results obtained for
different
values of $\kk$. Thus, for small $\kk$, the total energy of the system
can be expanded in powers of $\kk$
\begin{equation}
 \label{interactionA}
 E(\kk)=E_{0}+E_{1}\kk+E_{2}\kk^{2}+\sum_{k=3}
 ^\infty E_{k} \kk^k \,,
\end{equation}
\begin{equation}
 \label{div2}
 E_{k}=\frac{1}{k!}\frac{\rmd^k}{\rmd\kk^k}E(\kk)\Big|_{\kk=0}. 
\end{equation}
The higher-order contribution $E_{\geqslant 3}\equiv\sum_{k=3}^{\infty}E_k$
is calculated as $$E_{\geqslant 3}=E(\kk=1)-E_{0}-E_{1}-E_{2},$$ where the terms
$E_0$,  $E_1$, and $E_2$ are determined numerically according to Eq.~(\ref{div2}).
\begin{table}
\caption{The third- and higher-order interelectronic-interaction contributions
to the  $2p_{1/2}$-$2s$ and $2p_{3/2}$-$2s$ transition
energies in Li-like ions, in eV. The uncertainty due to the numerical procedure
is presented in the first brackets while the uncertainty due to the Breit
approximation is given in the second brackets.}
\linespread{1}
\label{table:E3}
\begin{center}
\begin{tabular}{rccllc}
 \hline
 \hline
$Z$&Interaction&Contribution&$2p_{1/2}-2s$&$2p_{3/2}-2s$&Ref.\\
 \hline
     3&C+B  &$E_3$&-0.4823  &-0.4839  &\\
     3&C+B  &$E_{\geqslant 3}$&-0.6483(20)(0)  &-0.6499(20)(0)  &\\
 \hline
     5&C+B  &$E_3$&-0.2860  &-0.2887  &\\
     5&C+B  &$E_{\geqslant 3}$&-0.3522(15)(0)  &-0.3548(15)(0)  &\\
 \hline
    10&C    &$E_3$&-0.1433  &-0.1466  &\\
    10&C    &$E_{\geqslant 3}$&-0.1583  &-0.1614  &\\
    10&C+B  &$E_3$&-0.1369  &-0.1423  &\\
    10&C+B  &$E_{\geqslant 3}$&-0.1545(6)(0)  &-0.1598(6)(0)  &\\
 \hline
    15&C+B  &$E_3$&-0.0858  &-0.0938  &\\
    15&C+B  &$E_{\geqslant 3}$&-0.0942(3)(0)  &-0.1025(3)(0)  &\\
 \hline
    20&C+B  &$E_3$&-0.0606  &-0.0719  &\\
    20&C+B  &$E_3$&-0.065   &         &\cite{Zherebtsov:2009}\\
    20&C+B  &$E_3$&-0.069   &  &\cite{Zherebtsov:00:227}\\
    20&C+B  &$E_{\geqslant 3}$&-0.0635(3)(0)  &-0.0747(3)(0)  &\\
    20&C+B  &$E_{\geqslant 3}$&-0.070   &         &\cite{Zherebtsov:00:227}\\
 \hline
    30&C    &$E_3$&-0.0406  &-0.0511  &\\
    30&C    &$E_3$&-0.045   &  &\cite{Zherebtsov:00:227}\\
    30&C    &$E_{\geqslant 3}$&-0.0418  &-0.0518  &\\
    30&C    &$E_{\geqslant 3}$&-0.046   &   &\cite{Zherebtsov:00:227}\\
    30&C+B  &$E_3$&-0.0284  &-0.0470  &\\
    30&C+B  &$E_3$&-0.0276  &-0.0463  &\cite{Yerokhin:pra:2007}\\
    30&C+B  &$E_3$&-0.060(8)  &   &\cite{Andreev:pra:2003}\\
    30&C+B  &$E_3$&-0.030$^{\ast}$  &   &\cite{Andreev:pra:2003}\\
    30&C+B  &$E_3$&-0.036   &   &\cite{Zherebtsov:00:227}\\
    30&C+B  &$E_{\geqslant 3}$&-0.0296(3)(1)  &-0.0481(3)(1)  &\\
    30&C+B  &$E_{\geqslant 3}$&-0.036   &   &\cite{Zherebtsov:00:227}\\
 \hline
\end{tabular}
\end{center}
\end{table}
\addtocounter{table}{-1}
\begin{table}
\caption{(Continued.)}
\linespread{1}
\begin{center}
\begin{tabular}{rlcllc}
 \hline
 \hline
$Z$&Interaction&Contribution&$2p_{1/2}-2s$&$2p_{3/2}-2s$&Ref.\\
 \hline
    35&C+B  &$E_3$&-0.0173  &-0.0401  &\\
    35&C+B  &$E_{\geqslant 3}$&-0.0181(3)(5)  &-0.0403(3)(5)  &\\
 \hline
    40&C+B  &$E_3$&-0.0070  &-0.0344  &\\
    40&C+B  &$E_3$&-0.009  &  &\cite{Zherebtsov:2009}\\
    40&C+B  &$E_3$&-0.015  &  &\cite{Zherebtsov:00:227}\\
    40&C+B  &$E_{\geqslant 3}$&-0.0077(4)(10)  &-0.0348(4)(10)  &\\
    40&C+B  &$E_{\geqslant 3}$&-0.015  &  &\cite{Zherebtsov:00:227}\\
 \hline
    45&C+B  &$E_3$& 0.0043  &-0.0286  &\\
    45&C+B  &$E_{\geqslant 3}$& 0.0017(6)(15)  &-0.0314(6)(15)  &\\
 \hline
    50&C    &$E_3$&-0.0120  &-0.0333  &\\
    50&C    &$E_3$&-0.014   &  &\cite{Zherebtsov:2009}\\
    50&C    &$E_3$&-0.016   &  &\cite{Zherebtsov:00:227}\\
    50&C    &$E_{\geqslant 3}$&-0.0133  &-0.0340  &\\
    50&C+B  &$E_3$& 0.0136  &-0.0271  &\\
    50&C+B  &$E_3$& 0.011  &  &\cite{Zherebtsov:2009}\\
    50&C+B  &$E_3$& 0.004  &  &\cite{Zherebtsov:00:227}\\
    50&C+B  &$E_{\geqslant 3}$& 0.0113(7)(20)  &-0.0283(7)(20)  &\\
 \hline
    54&C+B  &$E_3$& 0.0214  &-0.0250  &\\
    54&C+B  &$E_3$& 0.020  & &\cite{Zherebtsov:2009}\\
    54&C+B  &$E_3$& 0.012  & &\cite{Zherebtsov:00:227}\\
    54&C+B  &$E_{\geqslant 3}$& 0.0195(8)(25)  &-0.0260(8)(25)  &\\
 \hline
    60&C+B  &$E_3$& 0.0329  &-0.0236  &\\
    60&C+B  &$E_3$& 0.033   &  &\cite{Zherebtsov:2009}\\
    60&C+B  &$E_3$& 0.024   &  &\cite{Zherebtsov:00:227}\\
    60&C+B  &$E_3$& 0.043   &  &\cite{Andreev:pra:2003}\\
    60&C+B  &$E_{\geqslant 3}$& 0.0322(10)(30)  &-0.0239(10)(30)  &\\
 \hline
\end{tabular}
\end{center}
\end{table}
\addtocounter{table}{-1}
\begin{table}
\caption{(Continued.)}
\linespread{1}
\begin{center}
\begin{tabular}{rlcllc}
 \hline
 \hline
$Z$&Interaction&Contribution&$2p_{1/2}-2s$&$2p_{3/2}-2s$&Ref.\\
 \hline
    70&C+B  &$E_3$& 0.055  &-0.025  &\\
    70&C+B  &$E_3$& 0.047   &  &\cite{Zherebtsov:00:227}\\
    70&C+B  &$E_3$& 0.059(9)   &         &\cite{Andreev:pra:2003}\\
    70&C+B  &$E_3$& 0.052$^{\ast}$   &         &\cite{Andreev:pra:2003}\\
    70&C+B  &$E_{\geqslant 3}$& 0.054(2)(10)  &-0.024(2)(10)  &\\
 \hline
    80&C+B  &$E_3$& 0.084  &-0.029  &\\
    80&C+B  &$E_3$& 0.095   &  &\cite{Zherebtsov:2009}\\
    80&C+B  &$E_3$& 0.076   &  &\cite{Zherebtsov:00:227}\\
    80&C+B  &$E_3$& 0.099(14)   &  &\cite{Andreev:pra:2003}\\
    80&C+B  &$E_3$& 0.089$^{\ast}$   &  &\cite{Andreev:pra:2003}\\
    80&C+B  &$E_{\geqslant 3}$& 0.084(4)(13)  &-0.028(4)(13)  &\\
 \hline
    83&C    &$E_3$& 0.0328  &-0.0261  &\\
    83&C    &$E_3$& 0.031   &  &\cite{Zherebtsov:2009}\\
    83&C    &$E_3$& 0.029   &  &\cite{Zherebtsov:00:227}\\
    83&C    &$E_3$& 0.041   &-0.024   &\cite{Sapirstein:2001:022502}\\
83&C    &$E_{\geqslant 3}$& 0.0312  &-0.0275  &\\
    83&C+B  &$E_3$& 0.098   &-0.030  &\\
    83&C+B  &$E_3$& 0.103   &-0.019  &\cite{Yerokhin:pra:2007}\\
    83&C+B  &$E_3$& 0.104   &  &\cite{Zherebtsov:2009}\\
    83&C+B  &$E_3$& 0.087   &  &\cite{Zherebtsov:00:227}\\
    83&C+B  &$E_{\geqslant 3}$& 0.097(5)(15)  &-0.029(5)(15)  &\\
 \hline
    90&C+B  &$E_3$& 0.127  &-0.036  &\\
    90&C+B  &$E_3$& 0.147  &  &\cite{Zherebtsov:2009}\\
    90&C+B  &$E_3$& 0.118  &  &\cite{Zherebtsov:00:227}\\
    90&C+B  &$E_{\geqslant 3}$& 0.127(6)(40)  &-0.035(6)(40)  &\\
 \hline
 \hline
\end{tabular}
\end{center}
\end{table}
\addtocounter{table}{-1}
\begin{table}
\caption{(Continued.)}
\linespread{1}
\begin{center}
\begin{tabular}{rlcllc}
 \hline
 \hline
$Z$&Interaction&Contribution&$2p_{1/2}-2s$&$2p_{3/2}-2s$&Ref.\\
 \hline
    92&C+B  &$E_3$& 0.137  &-0.041  &\\
    92&C+B  &$E_3$& 0.160   &  &\cite{Zherebtsov:2009}\\
    92&C+B  &$E_3$& 0.131   &  &\cite{Zherebtsov:00:227}\\
    92&C+B  &$E_3$& 0.167(23)   &  &\cite{Andreev:pra:2003}\\
    92&C+B  &$E_3$& 0.147$^{\ast}$   &  &\cite{Andreev:pra:2003}\\
    92&C+B  &$E_{\geqslant 3}$& 0.137(7)(50)  &-0.039(7)(50)  &\\
 \hline
 \hline
\multicolumn{6}{l}{$^{\ast}$ The results of Ref. \cite{Andreev:pra:2003} with
the two Breit and one}\\
\multicolumn{6}{l}{\quad Coulomb photon-exchange contributions subtracted.}\\

\end{tabular}
\end{center}
\end{table}

The results of the numerical calculation of the higher-order 
interelectronic-interaction contributions for the $2p_{1/2}$-$2s$ and
$2p_{3/2}$-$2s$
transition energies in Li-like ions are collected in Table~\ref{table:E3}. 
``C'' in the second column indicates that only Coulomb interaction is taken into
account, while ``C+B'' means that both Coulomb and Breit interactions are
included. As one
can see from the table, in accordance with
Refs.~\cite{Zherebtsov:00:227,Yerokhin:pra:2007},
the Breit interaction contribution is rather significant,
especially for middle and high-$Z$ ions. We note that
the third-order contribution monotonously increases and changes the sign
when $Z$ increases. 
The uncertainty of the results consists of two parts: an uncertainty due to
some approximations made in the numerical
procedure, in the table it is written in the first brackets, and an uncertainty
due to the Breit approximation, it is given in the second
brackets.
To estimate the first uncertainty we studied
the convergence of the calculation depending on the configuration basis set
and compared our results with very accurate data obtained for lithium with the
variational solution of the three-body Schr\"{o}dinger problem
that includes the relativistic corrections obtained within the $\az$ expansion
\cite{Yan:prl+erratum:2008,Puchalski:pra:2006,Yan:02:042504}. 
The estimation of the residual three- and more photon-exchange QED effects is
more difficult.
As was found in Refs.~\cite{Yerokhin:pra:2001,Yerokhin:pra:2007} the QED
part
of the two-photon exchange correction is anomalously small for the $2s$ and $2p_{1/2}$
states. Moreover, the third order of the electron-electron interaction changes
its
sign when $Z$ increases. Thus, the value based on the ratio of the two-photon
exchange QED correction to corresponding non-QED contribution might
underestimate the three-photon QED
effects. 
For this reason, to estimate the uncertainty due to the QED effects, we take the
ratio of the QED and non-QED two-photon contributions for the
$2p_{3/2}-2s$ transition, where the QED effect is adequate, and multiply it by
the
maximal value of the third-order  contribution among the
$2s$, $2p_{1/2}$ and $2p_{3/2}$ states.

Comparing the results for the third and higher orders ($E_{\geqslant 3}$)
with the third order ($E_3$), we conclude that corrections of the fourth and
higher orders
($E_{\geqslant 3}-E_3$) are rather important, especially for low- and
middle-$Z$ ions.

We observe a reasonable agreement with Zherebtsov \textit{et al.}
\cite{Zherebtsov:00:227} and Yerokhin \textit{et al.} \cite{Yerokhin:pra:2007}.
A small discrepancy with the results of Yerokhin \textit{et al.} is caused by a
different way of taking into account the Breit interaction. Yerokhin
\textit{et al.}
treated the Breit interaction to first order only
(exchange by
only one Breit and two Coulomb photons), whereas we calculated so called
``iterated'' Breit interaction (exchange
by two
Breit and one Coulomb photons, and by three Breit photons). It should be
also mentioned
that Yerokhin \textit{et al.} \cite{Yerokhin:pra:2007} included the
negative-energy contribution for
the correction considered. However, this contribution is relatively
small.
As comparing to Andreev \textit{et al.}~\cite{Andreev:pra:2003}, a distinct
deviation is found. Most probably, as indicated in
Ref.~\cite{Yerokhin:pra:2007},
it is due to an overestimation of the contribution induced by
two Breit and one Coulomb
photon exchange in Ref.~\cite{Andreev:pra:2003}. The results with
this
contribution subtracted are marked by an asterisk in the table. They are
much closer to our results.

\begin{table}
\caption{Electronic-structure contributions to the $2p_{1/2}-2s$ and
$2p_{3/2}-2s$ transition energies in Li-like ions, in eV. The nuclear-charge
rms radii $\la r^2 \ra^{1/2}$ (in fm) are taken from
Refs.~\cite{ADNDT87_185,Kozhedub:pra:2008}.
The uncertainty given is due to the higher-order interelectronic interaction
only. The first one is caused by the numerical procedure while the second one is
due to the Breit approximation.
}
\label{Electron_structure}
\begin{center}
\begin{tabular}{lrrrrrrll}
 \hline
 \hline
$Z$& $\la r^2 \ra^{1/2}$&Transition& Dirac & 1ph&  2ph &$\geqslant $3ph&
Total &Total Ref. \cite{Yerokhin:pra:2007} \\
\hline
      3&2.431&$2p_{1/2}-2s$&   0.00000&  5.77750& -3.28214& -0.6484&  
1.8470(20)(0)&  1.8466(105)\\
      &&&&&&&&  $1.84812^\mathrm{a}$\\
      &&&&&&&&  $1.8486^\mathrm{b}$\\
      3&2.431& $2p_{3/2}-2s$&   0.00367&  5.76900& -3.27566& -0.6499&  
1.8471(20)(0)&1.8466(105)\\
      &&&&&&&&$1.84816^\mathrm{a}$\\
      &&&&&&&&  $1.8486^\mathrm{b}$\\
      5&2.406&$2p_{1/2}-2s$&   0.00000&  9.64178& -3.29242& -0.3522&  
5.9972(15)(0)& 5.9963(32)\\
      &&&&&&&&  $5.9986(3)^\mathrm{b}$\\
5&2.406& $2p_{3/2}-2s$&   0.02832&  9.60222& -3.27437& -0.3548&  
6.0014(15)(0)& 6.0004(32)\\
      &&&&&&&& $6.0027(3)^\mathrm{b}$\\
7&2.558&$2p_{1/2}-2s$&  -0.00001& 13.52503& -3.30788& -0.2354&  
9.9817(10)(0)& 9.9814(21)\\
      &&&&&&&& $9.9823(3)^\mathrm{b}$\\
7&2.558& $2p_{3/2}-2s$&   0.10889& 13.41639& -3.27242& -0.2391& 
10.0138(10)(0)&  10.0133(21)\\
      &&&&&&&& $10.0144(3)^\mathrm{b}$\\
10&3.005&$2p_{1/2}-2s$&  -0.00008& 19.40227& -3.34105& -0.1545& 
15.9067(6)(0)& 15.9064(10)\\
      &&&&&&&& $15.9068(3)^\mathrm{b}$\\
10&3.005& $2p_{3/2}-2s$&   0.45426& 19.08472& -3.26835& -0.1598& 
16.1108(6)(0)& 16.1105(10)\\
      &&&&&&&& $16.1111(3)^\mathrm{b}$\\
      15&3.189&$2p_{1/2}-2s$&  -0.00046& 29.40265& -3.42321& -0.0942& 
25.8848(3)(0)& 25.8851(5)\\
      &&&&&&&& $25.8848(3)^\mathrm{b}$\\
15&3.189& $2p_{3/2}-2s$&   2.30928& 28.32447& -3.25801& -0.1025& 
27.2732(3)(0)&  27.2734(5)\\
      &&&&&&&& $27.2735(3)^\mathrm{b}$\\
 \hline
 \hline
 \multicolumn{3}{l}{$^\mathrm{a}$ Reference \cite{Yan:02:042504}.}\\
 \multicolumn{3}{l}{$^\mathrm{b}$ Reference \cite{Johnson:88:2764}.}
\end{tabular}
\end{center}
\end{table}
\addtocounter{table}{-1}
\begin{table}
\caption{(Continued.)
}
\begin{center}
\begin{tabular}{lrrrrrrll}
 \hline
 \hline
$Z$& $\la r^2 \ra^{1/2}$&Transition& Dirac & 1ph&  2ph &$\geqslant $3ph&
Total&Total Ref. \cite{Yerokhin:pra:2007}\\
 \hline      
     18&3.427&$2p_{1/2}-2s$&  -0.00114& 35.57028& -3.48920& -0.0738& 
32.0061(3)(0)& 32.0060(5)\\
     18&3.427& $2p_{3/2}-2s$&   4.80429& 33.69830& -3.24944& -0.0838& 
35.1694(3)(0)&  35.1691(5)\\
     20&3.476&$2p_{1/2}-2s$&  -0.00185& 39.76906& -3.54046& -0.0635& 
36.1633(3)(0)& 36.1634(5)\\
     20&3.476& $2p_{3/2}-2s$&   7.34120& 37.19179& -3.24261& -0.0747& 
41.2157(3)(0)&  41.2155(5)\\
     21&3.544&$2p_{1/2}-2s$&  -0.00237& 41.89790& -3.56830& -0.0593& 
38.2679(3)(0)& 38.2682(5)\\
     21&3.544& $2p_{3/2}-2s$&   8.93553& 38.90849& -3.23885& -0.0712& 
44.5340(3)(0)&  44.5339(5)\\
     26&3.737&$2p_{1/2}-2s$&  -0.00676& 52.88124& -3.73042& -0.04110& 
49.1030(3)(0)& 49.1029(5)\\
     26&3.737& $2p_{3/2}-2s$&  21.16322& 47.14284& -3.21611& -0.0561& 
65.0339(3)(0)&  65.0333(5)\\
     28&3.775&$2p_{1/2}-2s$&  -0.00965& 57.45499& -3.80693& -0.0353& 
53.6031(3)(1)& 53.6034(5)\\
     28&3.775& $2p_{3/2}-2s$&  28.57046& 50.25007& -3.20517& -0.0518& 
75.5636(3)(1)&  75.5633(5)\\
     30&3.929&$2p_{1/2}-2s$&  -0.01434& 62.14676& -3.88932& -0.0296& 
58.2135(3)(1)& 58.2130(5)\\
     30&3.929& $2p_{3/2}-2s$&  37.79922& 53.23463& -3.19217& -0.0481& 
87.7936(3)(1)&  87.7926(5)\\
     36&4.188&$2p_{1/2}-2s$&  -0.03884& 77.03795& -4.18113& -0.0159& 
72.8021(3)(5)& 72.8013(6)\\
     36&4.188& $2p_{3/2}-2s$&  79.45643& 61.33404& -3.14609& -0.0390&
137.6054(3)(5)& 137.6044(6)\\
40&4.270&$2p_{1/2}-2s$&  -0.06839& 87.76278& -4.41614& -0.0077& 
83.2706(4)(10)& 83.2701(8)\\
     40&4.270& $2p_{3/2}-2s$& 122.39809& 65.89147& -3.10591& -0.0347&
185.1490(4)(10)& 185.1476(10)\\
     47&4.544&$2p_{1/2}-2s$&  -0.18121&108.43093& -4.91254&  0.0054&
103.3426(6)(15)& 103.3418(14)\\
     47&4.544& $2p_{3/2}-2s$& 238.40726& 71.83668& -3.01448& -0.0303&
307.1992(6)(15)& 307.1988(17)\\
     50&4.654&$2p_{1/2}-2s$&  -0.26811&118.16524& -5.16445&  0.0113&
112.7440(7)(20)& 112.7433(16)\\
     50&4.654& $2p_{3/2}-2s$& 308.58586& 73.43976& -2.96522& -0.0283&
379.0321(7)(20)& 379.0323(19)\\
     52&4.743&$2p_{1/2}-2s$&  -0.34806&124.99085& -5.34549&  0.0154&
119.3127(8)(22)& 119.3110(16)\\
     52&4.743& $2p_{3/2}-2s$& 363.69747& 74.14337& -2.92756& -0.0271&
434.8862(8)(22)& 434.8850(20)\\
     54&4.787&$2p_{1/2}-2s$&  -0.44234&132.10903& -5.53908&  0.0195&
126.1471(8)(28)&
126.1444(20)\\
     54&4.787& $2p_{3/2}-2s$& 426.27988& 74.52795& -2.88604& -0.0260&
497.8958(8)(28)& 497.8940(24)\\
60&4.912&$2p_{1/2}-2s$&  -0.88944&155.44266& -6.20420&  0.0322&
148.3812(10)(40)& 148.3786(25)\\
     60&4.912& $2p_{3/2}-2s$& 666.61398& 73.51308& -2.73855& -0.0239&
737.3646(10)(40)& 737.3621(35)\\ 
\hline
 \hline
\end{tabular}
\end{center}
\end{table}

\addtocounter{table}{-1}
\begin{table}
\caption{(Continued.)
}
\begin{center}
\begin{tabular}{lrrrrrrll}
 \hline
 \hline
$Z$& $\la r^2 \ra^{1/2}$&Transition& Dirac & 1ph&  2ph &$\geqslant $3ph&
Total&Total Ref. \cite{Yerokhin:pra:2007}\\
 \hline      
70&5.311&$2p_{1/2}-2s$&  -2.91444&202.61211& -7.64776&  0.0544&
192.104(2)(10) &192.1023(38)\\
     70&5.311& $2p_{3/2}-2s$&1299.24227& 62.72945& -2.38263&
-0.0242&1359.565(2)(10) &1359.5629(52)\\
     80&5.463&$2p_{1/2}-2s$&  -8.57680&264.30462& -9.68045&  0.0837&
246.131(4)(13)&
246.130(6)\\
     80&5.463& $2p_{3/2}-2s$&2359.15998& 36.01281& -1.82616&
-0.0279&2393.319(4)(13)&2393.317(8)\\
     83&5.521&$2p_{1/2}-2s$& -11.89801&286.67896&-10.44836&  0.0970&
264.430(5)(15)&
264.427(7)\\
     83&5.521& $2p_{3/2}-2s$&2792.20782& 23.81784& -1.60493&
-0.0290&2814.391(5)(15) &2814.392(9)\\
     90&5.710&$2p_{1/2}-2s$& -26.00449&348.27283&-12.62608&  0.1270&
309.769(6)(40)&
309.780(10)\\
     90&5.710& $2p_{3/2}-2s$&4077.38297&-14.49611& -0.95621&
-0.0350&4061.896(6)(40) &4061.908(11)\\

92&5.857&$2p_{1/2}-2s$& -33.304&368.83426&-13.37086&  0.1370&
322.296(7)(50)&
322.292(11)$^{\ast}$\\
     92&5.857& $2p_{3/2}-2s$&4527.933&-28.41302& -0.72818&
-0.0390&4498.753(7)(50) &4498.750(12)$^{\ast}$\\
 \hline
 \hline
  \multicolumn{9}{l}{$^{\ast}$ Corrected for the nuclear
deformation effect and the rms value from Ref. \cite{Kozhedub:pra:2008}.}\\
\end{tabular}
\end{center}
\end{table}

In Table~\ref{Electron_structure} we collect all the electronic-structure
contributions to the
transition energies and compare our results with those by other
authors. For comparison we chose the most recent data from
Ref.~\cite{Yerokhin:pra:2007}, which are in reasonable agreement with
others calculations.
Only for light ions with small $Z=3-15$, where the correlation effects are large
compared to the relativistic contributions, results of other works
(without
QED effects) are also presented.
The column labeled ``Dirac'' contains the energy value obtained from the Dirac
equation with an extended nucleus. The Fermi nuclear charge distribution
was employed. 
Except for uranium, the root-mean-square (rms) radii were taken
from Ref.~\cite{ADNDT87_185}. In case of uranium, we use
the rms value from Ref.~\cite{Kozhedub:pra:2008} and take into account the
nuclear deformation effect
(see Ref.~\cite{Kozhedub:pra:2008} for details). 
The two-photon exchange correction is evaluated within the framework of QED,
following our previous investigations \cite{Yerokhin:pra:2001,Artemyev:pra:03}.
The uncertainty given is due to the higher-order
interelectronic interaction only.
In addition to a different treatment of the Breit interaction in the present
work and in Ref.~\cite{Yerokhin:pra:2007} (see the related discussion above),
we note some difference in evaluation of the QED part of the two-photon
exchange contribution. In our work it was calculated with the pure Coulomb
potential while in Ref.~\cite{Yerokhin:pra:2007} a local screening potential
was employed.
We remind also the reader that, in accordance with our definition of the
electronic-structure part, the values in Table~\ref{Electron_structure} are
given in the nonrecoil limit.


%
%
\section{Screened QED corrections}
\label{ScrQED}
%
%
The screened QED contribution $\Delta E_{\rm scrQED}$ incorporates
the screened SE $\Delta E_{\rm scrSE}$ and screened
VP $\Delta E_{\rm scrVP}$ corrections. As concerns
the QED part of the two-photon exchange correction, it is included in the
electronic-structure contribution (see the previous section). Therefore, here we
restrict ourself with the contributions of the screened SE and
VP terms into the $2p_j - 2s$ transition energies
of Li-like ions.

First estimates of the screened QED corrections in
Li-like ions were performed in
Refs.~\cite{indelicato:1990:5139,Blundell:93:1790,Lindgren:pra:1993,
ynnerman:1994:4671,Chen:pra:1995}, where these
corrections were included either phenomenologically
or partly. The rigorous evaluations of the screened
SE and VP corrections were first performed in works
\cite{yerokhin:1999:3522,yerokhin:2005:12} and \cite{artemyev:1999:45},
respectively. These calculations incorporate the second-order QED
effects starting with the pure Coulomb potential as the zeroth-order
approximation (the original Furry picture). Later, in case of Li-like
bismuth, these corrections were calculated
starting with a local screening
potential (the extended Furry picture) \cite{Sapirstein:2001:022502}.

In the present paper the screened
SE and
VP corrections are evaluated within the extended Furry representation
for the ionization energies of the $2s$, $2p_{1/2}$, and $2p_{3/2}$ states
of Li-like ions in the range of the nuclear charge number $Z = 10 - 92$.
Employing the extended Furry representation, one partially takes into
account the higher-order electron-electron interaction effects, that are beyond
the
considered order of the perturbative expansion. This
approach can accelerate the convergence of the QED perturbation theory
with respect to the interelectronic-interaction effects, especially for
small values of $Z$, where the convergence of the perturbative
expansion becomes slower.

The Dirac equation in the extended Furry representation can be written as
\be
\label{Dirac}
  \Bigl[-i\balpha\cdot\bnabla + \beta + V_{\rm nuc}
        + V_{\rm scr}\Bigr] | n \ra = \veps_n | n \ra\,,
\ee
where $V_{\rm nuc}$ is the Coulomb potential of the extended nucleus and
$V_{\rm scr}$ is a local screening potential, which partially accounts for the
interaction between the valence electron and the closed core electrons.
We employ here the Kohn-Sham screening potential derived within the
density-functional theory \cite{kohn:1965:A1133},
\be
\label{Vscr}
  V_{\rm scr}(r) = \alpha \int_0^\infty \rmd r'\frac{1}{r_>}\rho_t(r')
   - \frac{2}{3}\frac{\alpha}{r}\left(
      \frac{81}{32\pi^2}r\rho_t(r)\right)^{1/3}\,.
\ee
This potential was successfully utilized in our previous QED calculations
for
the g factor and hyperfine splitting of Li-like ions
\cite{glazov:2006:330,oreshkina:2007:889,volotka:2008:062507}.
Here, $\rho_t$ denotes the total radial charge density distribution of
the core electrons ($b$) and the valence electron ($a$) 
\be
\label{rho_t}
  \rho_t(r) = \sum_b [G_b^2(r) + F_b^2(r)] + [G_a^2(r) + F_a^2(r)]\,,
  \hspace{1cm} \int_0^\infty \rmd r \, \rho_t(r) = n_b + 1\,,
\ee
where $n_b$ is the number of the core electrons. The Kohn-Sham
potential is constructed for the lithiumlike ground state, namely, for the
$(1s^2)2s$ state. In order to estimate the sensitivity of the result on
the choice of the potential we consider also the core-Hartree potential,
which is just a Coulomb potential generated by the core electrons.
The screening potentials are generated self-consistently by solving the
Dirac equation (\ref{Dirac}) until the energies of the core and valence
states become stable on the level of $10^{-9}$. The asymptotic behavior
of the Kohn-Sham potentials at large distances is restored by
introducing the Latter correction \cite{latter:1955:510}.

\begin{figure}
\includegraphics[width=0.8\textwidth]{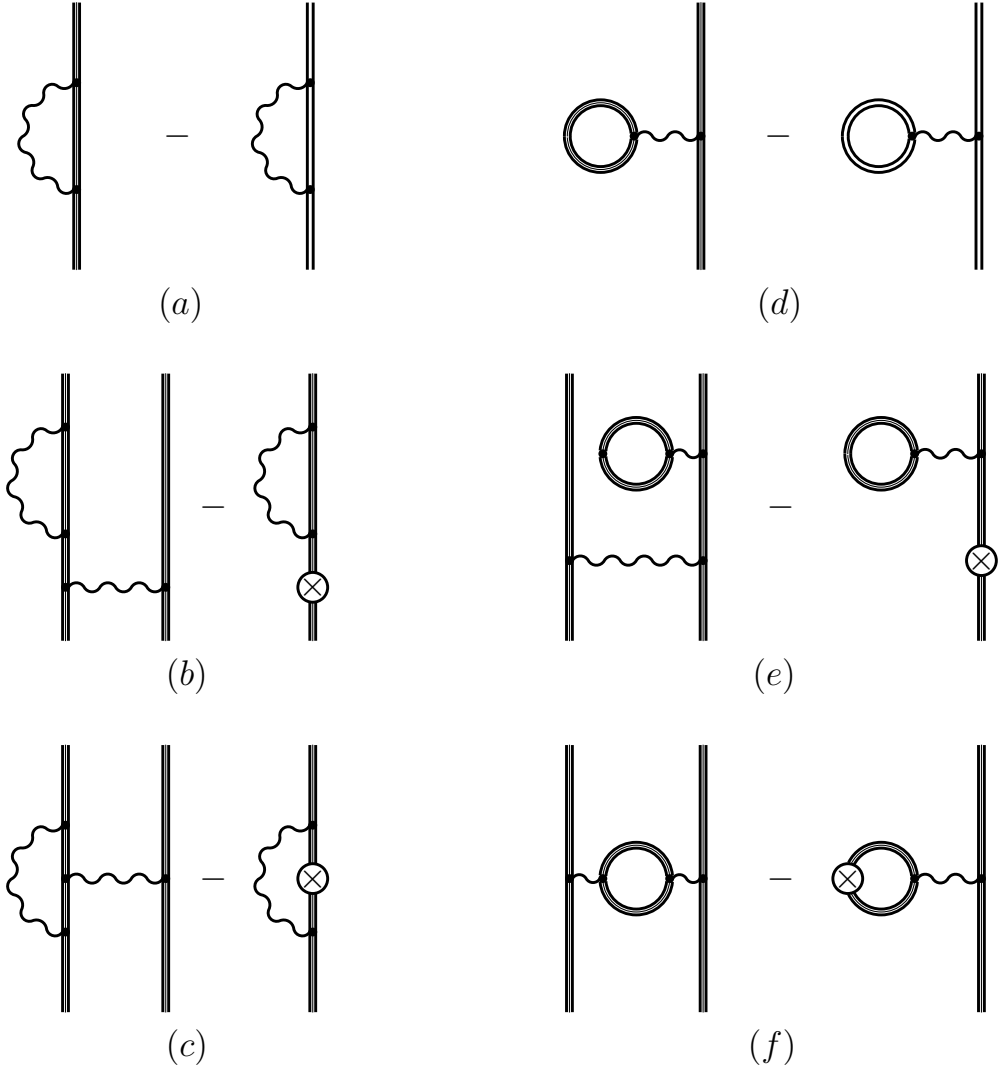}
\caption {Feynman diagrams representing the screened SE $(a-c)$
and VP $(d-f)$ corrections in the extended Furry
representation. The wavy line indicates the photon propagator, the
 triple line displays the electron propagators in the effective potential,
and the double line indicates the electron propagators in the Coulomb
field of the nucleus. The symbol $\otimes$ represents the extra
interaction term associated with the local screening potential.}
\label{fig:scr-eff}
\end{figure}

The complete gauge invariant set of diagrams which have to be considered
are shown in Fig.~\ref{fig:scr-eff}. They are referred to the SE
$(a-c)$ and VP $(d-f)$ diagrams.
The counterterm associated with the extra interaction term $V_{\rm scr}$
is represented graphically by the symbol $\otimes$. The formal
expressions for these diagrams are
derived from the first principles of QED employing the two-time
Green-function method
\cite{shabaev:2002:119}.

We consider here only the diagrams contributing to the ionization energy
of the valence state. It means that the one-electron core
and
core-core interaction diagrams are omitted in our consideration.
The corresponding contribution from the SE screening diagrams can
be written as
\be
  \Delta E_{\rm scrSE}^{(b)a} = \Delta E_{\rm scrSE}^{(0)} + \Delta E_{\rm scrSE}^{(1,\,\rm irr)}
               + \Delta E_{\rm scrSE}^{(1,\,\rm red)} + \Delta E_{\rm scrSE}^{(1,\,\rm ver)}\,.
\ee
The zero-order contribution $\Delta E_{\rm scrSE}^{(0)}$, depicted on
Fig.~\ref{fig:scr-eff}(a), is the difference between the SE
corrections calculated with and without the screening potential
\be
\label{se-0}
  \Delta E_{\rm scrSE}^{(0)} = \la a | \Sigma(\veps_a) | a \ra
             - \la a_{\rm C} | \Sigma(\veps_{a_{\rm C}}) | a_{\rm C} \ra\,.
\ee
Here, the subscript ``C'' labels the energies and wave functions
calculated with the Coulomb potential of the nucleus only, while
$\Sigma(\veps)$ denotes the unrenormalized self-energy operator.
The contribution of the diagrams depicted in Fig.~\ref{fig:scr-eff}(b)
is conveniently divided into irreducible and reducible parts
\cite{shabaev:2002:119}. The irreducible part is represented by the
expression
\be
\label{se-irr}
  \Delta E_{\rm scrSE}^{(1,\,\rm irr)} &=& 2 \sum_b \sum_P (-1)^P
   \Biggr[\sum_n^{\veps_n\neq\veps_a}
          \frac{\la Pa Pb | I(\Delta) | n b \ra \la n | \Sigma(\veps_a) | a \ra}{\veps_a-\veps_n}
\nonumber\\
      &+& \sum_n^{\veps_n\neq\veps_b}
          \frac{\la Pa Pb | I(\Delta) | a n \ra \la n | \Sigma(\veps_b) | b \ra}{\veps_b-\veps_n}
   \Biggl] - 2\sum_n^{\veps_n\neq\veps_a}
          \frac{\la a | V_{\rm scr} | n \ra \la n | \Sigma(\veps_a) | a \ra}{\veps_a-\veps_n}\,,
\ee
where the sum over $b$ runs over all core electron states, $P$ is the
permutation operator, giving rise to the sign $(-1)^P$ of the
permutation, $\Delta = \veps_{Pa} - \veps_a$, and $I(\omega)$ is the
interelectronic-interaction operator defined in the Coulomb gauge by
Eqs.~(\ref{I:def}) and (\ref{prop_C}). The expression for the reducible
part is given by
\be
\label{se-red}
  \Delta E_{\rm scrSE}^{(1,\,\rm red)} &=& \sum_b \sum_P (-1)^P
   \Biggr[\la Pa Pb | I(\Delta) | a b \ra
          \Bigr(\la a | \Sigma^\pr(\veps_a)| a \ra + \la b | \Sigma^\pr(\veps_b)| b \ra\Bigl)
\nonumber\\
      &-& \la Pa Pb | I^\pr(\Delta) | a b \ra
          \Bigr(\la a | \Sigma(\veps_a)| a \ra - \la b | \Sigma(\veps_b)| b \ra\Bigl)
   \Biggl] - \la a | V_{\rm scr} | a \ra \la a | \Sigma^\pr(\veps_a)| a \ra\,.
\ee
The vertex part, corresponding to Fig.~\ref{fig:scr-eff}(c), is given by
\be
\label{se-ver}
  \Delta E_{\rm scrSE}^{(1,\,\rm ver)} &=& \sum_b \sum_P (-1)^P
  \frac{i}{2\pi} \int^\infty_{-\infty}\rmd \omega \sum_{n_1 ,\, n_2}
  \Biggr[\frac{\la Pb n_1|I(\Delta)| b n_2 \ra \la Pa n_2|I(\omega)|n_1 a\ra}
        {(\veps_{Pa}-\omega-u\veps_{n_1})(\veps_a-\omega-u\veps_{n_2})}
\nonumber\\
     &+& \frac{\la Pa n_1|I(\Delta)| a n_2 \ra \la Pb n_2|I(\omega)|n_1 b\ra}
        {(\veps_{Pb}-\omega-u\veps_{n_1})(\veps_b-\omega-u\veps_{n_2})}
  \Biggl]
\nonumber\\
  &-& \frac{i}{2\pi} \int^\infty_{-\infty}\rmd \omega \sum_{n_1 ,\, n_2}
         \frac{\la n_1|V_{\rm scr}| n_2 \ra \la a n_2|I(\omega)|n_1 a\ra}
         {(\veps_a-\omega-u\veps_{n_1})(\veps_a-\omega-u\veps_{n_2})}\,,
\ee
where $u = 1 - i0$ preserves the proper treatment of poles of the
electron propagators. Expressions~(\ref{se-0})-(\ref{se-ver}) suffer
from ultraviolet divergences. To cancel these divergences explicitly
we have employed the renormalization scheme presented in details in
Refs.~\cite{yerokhin:1999:800,yerokhin:1999:3522}. The infrared
divergences which occur in some terms of the expressions (\ref{se-red}) and
(\ref{se-ver}) are regularized by introducing a nonzero photon mass and
canceled analytically.

The corresponding contributions of the screened VP
diagrams, depicted in Fig.~\ref{fig:scr-eff}(d)-(f), are
\be
  \Delta E_{\rm scrVP}^{(b)a} = \Delta E_{\rm scrVP}^{(0)} + \Delta E_{\rm scrVP}^{(1,\,\rm irr)}
               + \Delta E_{\rm scrVP}^{(1,\,\rm red)} + \Delta E_{\rm scrVP}^{(1,\,b)}\,,
\ee
\be
\label{vp-0}
  \Delta E_{\rm scrVP}^{(0)} = \la a | U_{\rm VP} | a \ra
             - \la a_{\rm C} | U_{\rm VP} | a_{\rm C} \ra\,,
\ee
\be
\label{vp-irr}
  \Delta E_{\rm scrVP}^{(1,\,\rm irr)} &=& 2 \sum_b \sum_P (-1)^P
   \Biggr[\sum_n^{\veps_n\neq\veps_a}
          \frac{\la Pa Pb | I(\Delta) | n b \ra \la n | U_{\rm VP} | a \ra}{\veps_a-\veps_n}
\nonumber\\
      &+& \sum_n^{\veps_n\neq\veps_b}
          \frac{\la Pa Pb | I(\Delta) | a n \ra \la n | U_{\rm VP} | b \ra}{\veps_b-\veps_n}
   \Biggl] - 2\sum_n^{\veps_n\neq\veps_a}
          \frac{\la a | V_{\rm scr} | n \ra \la n | U_{\rm VP} | a \ra}{\veps_a-\veps_n}\,,
\ee
\be
\label{vp-red}
  \Delta E_{\rm scrVP}^{(1,\,\rm red)} = -\sum_b \sum_P (-1)^P
   \la Pa Pb | I^\pr(\Delta) | a b \ra
   \Bigr(\la a | U_{\rm VP} | a \ra - \la b | U_{\rm VP} | b \ra\Bigl)\,,
\ee
\be
\label{vp-b}
  \Delta E_{\rm scrVP}^{(1,\,b)} = \sum_b \sum_P (-1)^P
   \la Pa Pb | I_{\rm VP}(\Delta) | a b \ra - \la a | U^{\rm scr}_{\rm VP} | a \ra\,,
\ee
where $U_{\rm VP}$ denotes the VP potential, and
$I_{\rm VP}(\Delta)$ is the interelectronic-interaction operator modified by the
electron-loop. For the renormalization
of the expressions~(\ref{vp-0})-(\ref{vp-b})
we refer to the works~\cite{soff:1988:5066,artemyev:1999:45}.
Accordingly, these contributions are divided into the Uehling and
Wichmann-Kroll parts. The renormalized Uehling parts of the
VP operators $U_{\rm VP}$ and $I_{\rm VP}(\Delta)$
are given by the expressions (see, e.g., Ref.~\cite{artemyev:1999:45})
\be
  U_{\rm VP}(r) &=& -\frac{2\alpha^2 Z}{3 r}\int_1^\infty
  \rmd t \, \frac{\sqrt{t^2-1}}{t^3} \left( 1+\frac{1}{2t^2} \right)
  \int_0^\infty \rmd r^\pr \, r^\pr \rho_{\rm eff}(r^\pr)
\nonumber\\
&\times& \left[\exp{(-2|r-r^\pr|t)}-\exp{(-2|r+r^\pr|t)}\right]\,,
\ee
\be
\label{I_VP-def}
  I_{\rm VP}(\Delta,r_{12}) = \alpha\,\frac{\alpha_{1\mu}\alpha_2^\mu}{r_{12}}\,
  \frac{2\alpha}{3\pi} \int_1^\infty
  \rmd t \, \frac{\sqrt{t^2-1}}{t^2} \left( 1+\frac{1}{2t^2} \right)
  \exp{(-\sqrt{4t^2-\Delta^2}\,r_{12})}\,,
\ee
where the density $\rho_{\rm eff}$ is related to the nuclear binding
and local screening potentials via the Poisson equation
$\Delta V_{\rm nuc}(r) + \Delta V_{\rm scr}(r)=4\pi \alpha Z \rho_{\rm eff}(r)$.
$U^{\rm scr}_{\rm VP}$ differs from $U_{\rm VP}$ only by replacing
$\rho_{\rm eff}$ with $\rho_{\rm scr}$, where the density $\rho_{\rm scr}$
is related to the screening potential $V_{\rm scr}$.
The Wichmann-Kroll
parts of the expressions (\ref{vp-0})-(\ref{vp-red})
are evaluated employing the approximate formula for the Wichmann-Kroll
potential
\cite{fainshtein:1990:559}. The 
Wichmann-Kroll contribution to Eq.~(\ref{vp-b}) is relatively small
\cite{artemyev:1999:45} and
is neglected in the present consideration.

The numerical evaluation is based on the wave functions constructed from
B-splines employing the dual-kinetic-balance finite basis set
method \cite{shabaev:2004:130405}. The sphere model for the nuclear charge
distribution is used together with the rms radii taken from
Ref.~\cite{ADNDT87_185}, with the exception of the uranium ion, for which
the rms value is taken from work~\cite{Kozhedub:pra:2008}.
The calculations have been performed in both Feynman and Coulomb gauges for
the photon propagator describing the electron-electron interaction.
The results agree very well with each other, thus providing an
accurate check of the numerical procedure. In Table~\ref{tab:comp}
we compare our values of the screened SE
and VP corrections, calculated in the Kohn-Sham, core-Hartree,
and Coulomb potentials (as zeroth-order approximation), with other theoretical
results. As one can see from the table, our values for the screened
SE and VP corrections in the Coulomb potential
are in perfect agreement with the corresponding results of works
\cite{yerokhin:1999:3522,yerokhin:2005:12} and \cite{artemyev:1999:45},
respectively. As to comparison with the related values from
Ref.~\cite{Sapirstein:2001:022502}, some deviation can be stated for
both screened SE and VP contributions.
This discrepancy is especially noticeable for the $2p_{1/2}$ and
$2p_{3/2}$
screened SE terms. The reason of this disagreement is unclear for us.

In Table~\ref{tab:total} we present our results for the
total screened QED correction to the ionization energies of the $2s$, $2p_{1/2}$,
and $2p_{3/2}$ valence states as well as to the energy differences $2p_j-2s$,
calculated in the Kohn-Sham potential.
The corresponding results obtained in the core-Hartree potential are rather
close to the Kohn-Sham ones. Therefore, for the conservative estimation of the
theoretical uncertainty of the ionization energies due to the higher-order
contributions we consider the difference between the values obtained in
the Coulomb 
and Kohn-Sham potentials and assign the uncertainty to be $30\%$ of this
difference.
The related uncertainty for the energy differences $2p_j-2s$ is determined to be
the maximum of the error bars for the $2p_j$ and $2s$ states.
%
%
\begin{table}
\caption{The contributions of the screened self-energy
$\Delta E^{(b)a}_{\rm scrSE}$ and screened vacuum-polarization
$\Delta E^{(b)a}_{\rm scrVP}$ corrections for the $2s$, $2p_{1/2}$,
and $2p_{3/2}$ states of Li-like ions for different starting potentials,
in eV. Comparison with the other theoretical calculations is given.}
\label{tab:comp}
\tabcolsep3mm
\begin{tabular}{lllllll} \hline
    & \multicolumn{2}{c}{Kohn-Sham}
       & \multicolumn{2}{c}{core-Hartree}
          & \multicolumn{2}{c}{Coulomb} \\
$Z$ & $\Delta E^{(b)a}_{\rm scrSE}$ & $\Delta E^{(b)a}_{\rm scrVP}$
       & $\Delta E^{(b)a}_{\rm scrSE}$ & $\Delta E^{(b)a}_{\rm scrVP}$
          & $\Delta E^{(b)a}_{\rm scrSE}$ & $\Delta E^{(b)a}_{\rm scrVP}$ \\ \hline
\multicolumn{7}{c}{$2s$ state}  \\
 20 & $-0.0444$ & $0.0030$
       & $-0.0443$ & $0.0030$
          & $-0.0462$ & $0.0032$     \\
    &        &       
       &        &       
          & $-0.04624(3)^a$ & $0.0032^b$     \\
 50 & $-0.4782$ & $0.0587$
       & $-0.4775$ & $0.0586$
          & $-0.4879$ & $0.0599$     \\
    &        &       
       &        &       
          & $-0.4881(3)^a$ & $0.0599^b$     \\
 83 & $-2.318$ & $0.494$
       & $-2.315$ & $0.494$
          & $-2.356$ & $0.503$   \\
    &        &       
       &        &       
          & $-2.3553(2)^a$ & $0.5034(3)^b$     \\
    & $-2.317^c$ & $0.516^c$
       & $-2.311^c$ & $0.523^c$
          & $-2.363^c$ & $0.527^c$ \\
\multicolumn{7}{c}{$2p_{1/2}$ state}  \\
 20 & $-0.0083$ & $0.0007$
       & $-0.0083$ & $0.0007$
          & $-0.0098$ & $0.0009$     \\
    &        &       
       &        &       
          & $-0.00983(10)^d$ & $0.0009^b$     \\
 50 & $-0.1240$ & $0.0186$
       & $-0.1239$ & $0.0186$
          & $-0.1341$ & $0.0199$     \\
    &        &       
       &        &       
          & $-0.1341(3)^d$ & $0.0200^b$     \\
 83 & $-1.069$ & $0.244$
       & $-1.065$ & $0.243$
          & $-1.123$ & $0.256$   \\
    &        &       
       &        &       
          & $-1.1218(12)^d$ & $0.2564(1)^b$     \\
    & $-1.120^c$ & $0.268^c$
       & $-1.102^c$ & $0.268^c$
          & $-1.168^c$ & $0.276^c$ \\
\multicolumn{7}{c}{$2p_{3/2}$ state}  \\
 20 & $-0.0126$ & $0.0007$
       & $-0.0126$ & $0.0007$
          & $-0.0145$ & $0.0008$     \\
    &        &       
       &        &       
          & $-0.01458(3)^a$ & $0.0008^b$     \\
 50 & $-0.1603$ & $0.0121$
       & $-0.1603$ & $0.0121$
          & $-0.1701$ & $0.0129$     \\
    &        &       
       &        &       
          & $-0.1702(3)^a$ & $0.0129^b$     \\
 83 & $-0.752$ & $0.069$
       & $-0.751$ & $0.069$
          & $-0.776$ & $0.072$   \\
    &        &       
       &        &       
          & $-0.7763(6)^a$ & $0.0719^b$     \\
    & $-0.748^c$ & $0.088^c$
       & $-0.737^c$ & $0.087^c$
          & $-0.816^c$ & $0.087^c$ \\ \hline
\end{tabular}
\newline
$^a$ Yerokhin {\it et al.} \cite{yerokhin:2005:12}.
\hspace{1cm}
$^c$ Sapirstein and Cheng \cite{Sapirstein:2001:022502}.\\
$^b$ Artemyev {\it et al.} \cite{artemyev:1999:45}.
\hspace{1cm}
$^d$ Yerokhin {\it et al.} \cite{yerokhin:1999:3522}.
\end{table}
%
%
\begin{table}
\caption{The screened QED contributions to the ionization energies of the $2s$, $2p_{1/2}$,
and $2p_{3/2}$ states and to the energy differences $2p_{1/2}-2s$ and $2p_{3/2}-2s$
in Li-like ions, in eV.}
\label{tab:total}
\tabcolsep3mm
\begin{tabular}{lllllll} \hline
$Z$ & $\la r^2 \ra^{1/2}$ & $2s$ & $2p_{1/2}$ & $2p_{3/2}$ & $2p_{1/2}-2s$ & $2p_{3/2}-2s$ \\ \hline
10 & 3.005 & $-$0.0070(2)  & $-$0.0012(1)  & $-$0.0017(2)  &  0.0058(2)  &  0.0053(2)      \\
12 & 3.057 & $-$0.0113(2)  & $-$0.0019(2)  & $-$0.0028(3)  &  0.0094(2)  &  0.0085(3)      \\
14 & 3.122 & $-$0.0168(3)  & $-$0.0029(2)  & $-$0.0044(3)  &  0.0138(3)  &  0.0123(3)      \\
15 & 3.189 & $-$0.0200(3)  & $-$0.0035(3)  & $-$0.0053(4)  &  0.0165(3)  &  0.0146(4)      \\
18 & 3.427 & $-$0.0317(4)  & $-$0.0057(4)  & $-$0.0089(5)  &  0.0260(4)  &  0.0228(5)      \\
20 & 3.476 & $-$0.0414(5)  & $-$0.0076(4)  & $-$0.0119(5)  &  0.0338(5)  &  0.0294(5)      \\
21 & 3.544 & $-$0.0467(6)  & $-$0.0087(4)  & $-$0.0137(5)  &  0.0380(6)  &  0.0330(6)      \\
26 & 3.737 & $-$0.0797(8)  & $-$0.0151(7)  & $-$0.0243(8)  &  0.0646(8)  &  0.0554(8)      \\
28 & 3.775 & $-$0.0959(9)  & $-$0.0184(8)  & $-$0.0296(10) &  0.0775(9)  &  0.0662(10)     \\
30 & 3.929 & $-$0.1139(10) & $-$0.0221(9)  & $-$0.0357(11) &  0.0917(10) &  0.0782(11)     \\
32 & 4.074 & $-$0.1339(11) & $-$0.0266(10) & $-$0.0426(12) &  0.1073(11) &  0.0913(12)     \\
36 & 4.188 & $-$0.1800(14) & $-$0.0372(13) & $-$0.0590(15) &  0.1428(14) &  0.1211(15)     \\
40 & 4.270 & $-$0.2351(17) & $-$0.0511(15) & $-$0.0790(18) &  0.1840(17) &  0.1561(18)     \\
47 & 4.544 & $-$0.3564(22) & $-$0.0856(22) & $-$0.1243(24) &  0.2708(22) &  0.2322(24)     \\
50 & 4.654 & $-$0.4195(26) & $-$0.1054(26) & $-$0.1482(27) &  0.3141(26) &  0.2713(27)     \\
52 & 4.735 & $-$0.4657(27) & $-$0.1210(28) & $-$0.1663(27) &  0.3447(28) &  0.2994(27)     \\
54 & 4.787 & $-$0.5154(30) & $-$0.1381(32) & $-$0.1851(31) &  0.3773(32) &  0.3303(31)     \\
60 & 4.912 & $-$0.6883(38) & $-$0.2042(42) & $-$0.2522(38) &  0.4841(42) &  0.4361(38)     \\
66 & 5.221 & $-$0.903(5)   & $-$0.298(5)   & $-$0.335(4)   &  0.604(5)   &  0.567(5)       \\
70 & 5.312 & $-$1.073(5)   & $-$0.381(7)   & $-$0.401(5)   &  0.692(7)   &  0.672(5)       \\
74 & 5.367 & $-$1.269(6)   & $-$0.485(8)   & $-$0.475(6)   &  0.784(8)   &  0.794(6)       \\
79 & 5.436 & $-$1.556(7)   & $-$0.652(10)  & $-$0.583(6)   &  0.904(10)  &  0.973(7)       \\
80 & 5.463 & $-$1.620(8)   & $-$0.692(11)  & $-$0.606(7)   &  0.928(11)  &  1.014(8)       \\
82 & 5.501 & $-$1.753(8)   & $-$0.778(12)  & $-$0.656(7)   &  0.976(12)  &  1.097(8)       \\
83 & 5.521 & $-$1.824(9)   & $-$0.824(13)  & $-$0.683(6)   &  1.000(13)  &  1.141(9)       \\
90 & 5.710 & $-$2.394(11)  & $-$1.241(17)  & $-$0.882(8)   &  1.153(17)  &  1.512(11)      \\
92 & 5.857 & $-$2.584(11)  & $-$1.394(19)  & $-$0.948(9)   &  1.190(19)  &  1.637(11)      \\ \hline
\end{tabular}
\end{table}
%


\section{$2p_{j}-2s$ transition energies in Li-like ions}

In this section, we collect all theoretical contributions available for the
$2p_{1/2}-2s$ and $2p_{3/2}-2s$ transition energies for middle-$Z$
Li-like ions,
compare them with experimental results, and discuss prospects for further
improvement of the theoretical accuracy. Individual contributions to the
$2p_{1/2}-2s$ and $2p_{3/2}-2s$ transition energies are presented in
Tables~\ref{Total_table1}
and \ref{Total_table2}, respectively. The rms radii and their uncertainties
are listed in the second column of the tables. These values are taken from
Ref.~\cite{ADNDT87_185}. The uncertainty of the electronic-structure values
includes an error due to the model-dependence
of the nuclear charge distribution. It is conservatively estimated by comparing
the
results obtained within the Fermi and the homogeneously charged-sphere model.
Except for neon $(Z=10)$, the electronic-structure contributions given are
obtained in this work. In case of neon, we use the related result of
Ref.~\cite{Johnson:88:2764}, which has a higher accuracy.

Next, one should take into account the first-order
one-electron QED corrections.
They are determined by the SE and the VP. The
SE correction is obtained by interpolating the values presented in
Ref.~\cite{Beier:pra:1998}
for the $2s$ and $2p_{1/2}$ states and in
Ref.~\cite{Mohr:pra:1992}
for the $2p_{3/2}$ state. 
The Uehling part of the VP contribution was calculated in the present work
while the Wichmann-Kroll part is taken from Ref.~\cite{Beier:jpb:1997}.

The next corrections, which caused
 the largest theoretical uncertainties for middle-$Z$ ions
\cite{Yerokhin:pra:2007}, are the nuclear
recoil and screened QED contributions. The recoil effect is
considered in Sec.~\ref{MS_sec}, while the evaluation of the screened QED
corrections is presented in Sec.~\ref{ScrQED}.
These calculations improve considerably the accuracy of the theoretical
predictions
for the $2p_j-2s$ transition energies in middle-$Z$ Li-like ions.

Finally, we should account for the two-loop one-electron QED effect. So-called
``SEVP'',
``VPVP'', and ``S(VP)E'' subsets were recently tabulated in
Ref.~\cite{Yerokhin:pra:2008}.
The remaining two-loop SE correction (the ``SESE'' subset) for $n=2$
states was accomplished only for several ions with $Z\geq60$
\cite{Yerokhin:prl:2006}. In order to
obtain the SESE correction for middle-$Z$ ions we use an
extrapolation procedure. For the $2s$ state, the extrapolation is performed
in two steps. At first, the numerical values for the $1s$ state are obtained by
interpolating the numerical results of
Refs.~\cite{Yerokhin:pra:2005,Yerokhin:prl:2006,Yerokhin:pra:2009}.
Then the weighted difference 
$\Delta_s=8\delta E_{2s}-\delta E_{1s}$ is achieved by using low-order
terms of the $\alpha Z$-expansion and extrapolating the
higher-order contributions from the all-order results
(see Ref.~\cite{Yerokhin:cjp:2007} and references therein).
An uncertainty of $30\%$ is assigned to these results.  For the
$2p_j$ states,
the correction is much smaller and, for our purpose, it is sufficient to use the
$\alpha Z$-expansion \cite{Yerokhin:cjp:2007} with the boundaries for the higher-order
remainder $\pm2\alpha^2(\alpha Z)^6/(8\pi^2)$.

As one can see from the tables, the total theoretical results agree well with
the experimental data. Compared to the experimental accuracy, the theoretical
one is generally better, almost the same in the cases of argon ($Z=18$) and iron
($Z=26$), and worse for neon ($Z=10$) and scandium ($Z=21$, the $2p_{3/2}-2s$
transition). 
For middle-$Z$ ions, the leading theoretical uncertainties  arise from
the higher-order screened QED and the
electronic-structure contributions. For $Z$ greater than 40 the uncertainty due
to the
two-loop one-electron QED corrections becomes also considerable. We conclude
that the 
present status of the theory and experiment for middle-$Z$ Li-like ions provides
a test of QED on a level of a few tenths of a percent. 

Further improvements of the theoretical predictions can be achieved by
calculating  the screened QED corrections of the second
order in $1/Z$ and 
the three-photon exchange QED corrections.

\begin{table}
\caption{Individual contributions to the $2p_{1/2}-2s$ transition energy in
Li-like ions, in eV.
}
\label{Total_table1}
\begin{tabular}{lllllllllllr}
\hline
 \hline
$Z$& $\la r^2\ra^{1/2}$&Electronic & 1-loop QED & Scr.QED & Recoil
&2-loop QED& Total& Experiment& Ref.\\ 
&&structure & & &&& theory&&\\ \hline
     10& 3.005(2)& 15.9068(3)& -0.0200&  0.0058(2)& -0.0042&  0.0000&
15.8883(4)&15.8887(2)&\cite{Edlen:ps:1983} \\
     15& 3.189(2)& 25.8848(3)& -0.0833&  0.0165(3)& -0.0071&  0.00005&
25.8110(4)&
25.814(3)&\cite{Martin:jpcrd:1985}\\
     18& 3.427(2)& 32.0061(3)& -0.1569&  0.0260(4)& -0.0081&  0.0001&
31.8673(5)&
     31.8664(9)&\cite{Edlen:ps:1983}
\\
     20& 3.476(1)& 36.1633(3)& -0.2260&  0.0338(5)& -0.0100&  0.0002&
35.9612(6)&
     35.9625(25)&\cite{Sugar:jpcrds:1985}
\\
     21& 3.544(2)& 38.2679(3)& -0.2673&  0.0380(6)& -0.0099&  0.00024&
38.0289(7)&
     38.02(4)&\cite{Suckewer:pl:1980}\\
     26& 3.737(2)& 49.1030(3)& -0.5565&  0.0646(8)& -0.0126&  0.0007(1)&
48.5991(9)& 48.5982(8)&\cite{Epp:prl:2007} \\
     &&&&&&&& 48.5997(10)&\cite{Reader:josab:1994} \\
     28& 3.775(1)& 53.6031(3)& -0.7169&  0.0775(9)& -0.0142&  0.0009(1)&
52.9504(10)& 52.9501(11)&\cite{Sugar:josab:1992,Sugar:josab:1993}\\
     30& 3.929(1)& 58.2135(3)& -0.9070&  0.0917(10)& -0.0149&  0.0013(2)&
57.3846(10)& 57.3839(30)&\cite{Staude:pra:1998}\\
     36& 4.188(1)& 72.8021(6)& -1.6859&  0.1428(14)& -0.0167&  0.0029(5)&
71.2451(15)& 71.243(8)&\cite{Madzunkov:pra:2002}\\
     &&&&&&&& 71.241(11)&\cite{Hinnov:pra:1989} \\
40& 4.270(1)& 83.2706(11)& -2.4107&  0.1840(17)& -0.0195&  0.0046(10)&
81.0289(23)& \\
     47& 4.544(4)&103.3426(17)& -4.1673&  0.2708(22)& -0.0233&  0.0094(21)&
99.4321(35)& 99.438(7)&\cite{Bosselmann:pra:1999}\\
     50& 4.654(1)&112.7440(22)& -5.1431&  0.3141(26)& -0.0238& 
0.0124(30)&107.9036(45)& 107.911(7)&\cite{Feili:pra:2000}\\
     52& 4.743(3)&119.3127(24)& -5.8777&  0.3447(28)& -0.0243& 
0.0147(35)&113.770(5)& \\
     54& 4.787(5)&126.1471(31)& -6.6851&  0.3773(32)& -0.0257& 
0.0175(40)&119.831(6)& 119.820(8)&\cite{Feili:pra:2000}\\
     60& 4.912(2)&148.3812(40)& -9.5873&  0.4841(42)& -0.0305& 
0.0271(15)&139.275(6)& \\
\hline
 \hline
\end{tabular}
\end{table}

\begin{table}
\caption{Individual contributions to the $2p_{3/2}-2s$ transition energy in
Li-like ions, in eV. 
}
\label{Total_table2}
\begin{tabular}{lllllllllllr}
 \hline
 \hline
$Z$& $\la r^2\ra^{1/2}$&Electronic & 1-loop QED & Scr.QED & Recoil
&2-loop QED& Total& Experiment& Ref.\\ 
&&structure & & &&& theory&&\\ \hline
     10& 3.005(2)& 16.1111(3)& -0.0190&  0.0053(2)& -0.0042&  0.0000&
16.0932(4)&
16.0932(2)&\cite{Edlen:ps:1983}\\
     15& 3.189(2)& 27.2732(3)& -0.0781&  0.0146(4)& -0.0071&  0.00004&
27.2026(5)& 27.206(3)&\cite{Martin:jpcrd:1985}\\
      18& 3.427(2)& 35.1694(3)& -0.1463&  0.0228(5)& -0.0082&  0.0001&
35.0378(6)&
     35.0370(12)&\cite{Edlen:ps:1983}
\\
20& 3.476(1)& 41.2157(3)& -0.2100&  0.0294(5)& -0.0101&  0.0002&
41.0251(7)& 
     41.0286(25)&\cite{Sugar:jpcrds:1985}\\
    21& 3.544(2)& 44.5340(3)& -0.2480&  0.0330(6)& -0.0100&  0.0002&
44.3092(7)&
     44.3094(2)&\cite{Lestinsky:prl:2008}\\
     26& 3.737(2)& 65.0339(3)& -0.5119&  0.0554(8)& -0.0128&  0.0005(1)&
64.5650(9)& 64.5657(17)&\cite{Reader:josab:1994} \\
     28& 3.775(1)& 75.5636(3)& -0.6574&  0.0662(10)& -0.0145&  0.0008(1)&
74.9586(11)& 74.9602(22)&\cite{Sugar:josab:1992,Sugar:josab:1993} \\
     30& 3.929(1)& 87.7936(3)& -0.8294&  0.0782(11)& -0.0153&  0.0011(2)&
87.0282(12)&  87.0302(37)&\cite{Staude:pra:1998}\\
     36& 4.188(1)&137.6054(6)& -1.5298&  0.1211(15)& -0.0173& 
0.0024(5)&136.1818(17)&
136.202&\cite{nist}\\
     &&&&&&&& 136.173(37)&\cite{Hinnov:pra:1989} \\
     40& 4.270(1)&185.1490(11)& -2.1781&  0.1561(18)& -0.0204& 
0.0039(10)&183.1106(23)& \\
     47& 4.544(4)&307.1992(16)& -3.7439&  0.2322(24)& -0.0247& 
0.0081(21)&303.6709(36)&303.67(3)&\cite{Bosselmann:pra:1999}\\ 
     50& 4.654(1)&379.0321(22)& -4.6129&  0.2713(27)& -0.0255& 
0.0107(30)&374.6757(46)& \\
     52& 4.743(3)&434.8862(24)& -5.2677&  0.2994(27)& -0.0262& 
0.0128(35)&429.904(5)& \\
     54& 4.787(5)&497.8958(31)& -5.9880&  0.3303(31)& -0.0278& 
0.0152(40)&492.225(6)& 492.34(62)&\cite{Martin:el:1989}\\
     60& 4.912(2)&737.3646(40)& -8.5884&  0.4361(38)& -0.0338& 
0.0253(20)&729.204(6)& \\
 \hline
 \hline
\end{tabular}
\end{table}

%
\section{Conclusion}

We have presented a systematic evaluation of the relativistic nuclear recoil
effect in Li-like ions. The recoil correction within the leading relativistic
approximation was
calculated with many-electron wave functions in order to take into account the
electron correlation effect. It relies on the large-scale CI-DFS method.
The higher-order relativistic recoil correction were also taken into account.
The results obtained are used to evaluate the $2p_j-2s$ transition energies.
They can also be employed to get the isotope
shifts in Li-like ions.

A systematic QED treatment of the electron correlation for the $2p_j-2s$
transitions in Li-like ions was presented. The rigorous QED calculation of the
one-
and two-photon exchange contributions  is combined with the
electron correlations of third and higher orders, that have been 
evaluated within the Breit approximation employing the CI-DFS method. The
complete gauge invariant sets of the screened one-loop QED corrections
have been rigorously evaluated. Different local potentials were used as the
zeroth-order approximation, namely, the Coulomb, core-Hartree, and
Kohn-Sham potentials.
The screened QED
contributions to the ionization energies of the $2s$, $2p_{1/2}$, and $2p_{3/2}$
states
as well as to the $2p_j-2s$ transition energies are presented for Li-like ions 
in the range $Z = 10 - 92$.

Finally, we have compiled all available theoretical contributions to the
$2p_j-2s$ transition energies in middle-$Z$ Li-like ions for $Z = 10 - 60$.
Due to the more elaborative evaluations of the electron-electron interaction in
the relativistic recoil
and QED contributions we have substantially reduced the total uncertainty of the
theoretical predictions. A good agreement with the experimental results has been
found.
%
\section{Acknowledgments}
We thank O. Zherebtsov for providing us with his unpublished results.
Valuable communications with V. Yerokhin are gratefully acknowledged.
The authors acknowledge the support by RFBR (Grant No. 07-02-00126-a),
GSI, DFG (Grant No. 436RUS113/950/0-1), and by the Ministry of Education
and Science of Russian Federation (Program for Development of Scientific
Potential of High School, Grant No. 2.1.1/1136; Program "Scientific and
pedagogical specialists for innovative Russia", Grant No. P1334). 
Y.S.K. acknowledges support by the Dynasty Foundation and DAAD.
The work of A.N.A. is supported by the Helmholtz Gemeinschaft and GSI under the
project VH--NG--421. 
The work of D.A.G. is
supported by the grant of the President of Russian
Federation, the Saint-Petersburg Government, and the
FAIR--Russia Research Center.
V.M.S. acknowledges the
support by the Alexander von Humboldt Foundation.
\input{Li-like2.bbl}
\end{document}